\documentclass[twocolumn]{aastex631}

\usepackage{amsmath}
\usepackage{hyperref}
\usepackage{makecell}
\usepackage{booktabs}    

\begin{document}

\title{Fermi-LAT View on Three Ultra-high-energy 1LHAASO Sources in the $52^{\circ}<l<55^{\circ}$ Region}


\author[0009-0004-6750-821X]{Linjie Liu}
\affiliation{Yunnan Observatories, Chinese Academy of Sciences, Kunming 650216, P. R. China}
\affiliation{University of Chinese Academy of Sciences, Beijing 100049, P. R. China}
\author[0000-0003-0933-6101]{Xian Hou}
\affiliation{Yunnan Observatories, Chinese Academy of Sciences, Kunming 650216, P. R. China}
\author{Pierrick Martin}
\affiliation{IRAP, Universit\'e de Toulouse, CNRS, UPS, CNES, F-31028 Toulouse, France}
\author{Chuyuan Yang}
\affiliation{Yunnan Observatories, Chinese Academy of Sciences, Kunming 650216, P. R. China}
\email{xhou@ynao.ac.cn}
\email{pierrick.martin@irap.omp.eu}

\begin{abstract}

Using more than 17 yr of \textit{Fermi}-LAT data, we performed a detailed investigation of the complex $52^{\circ}<l<55^{\circ}$ region, which encompasses the three ultra-high-energy sources 1LHAASO J1928+1746u, 1LHAASO J1928+1813u, and 1LHAASO J1929+1846u.
This region hosts multiple supernova remnants (SNRs), pulsars, GeV and TeV sources. Our analysis resolves the GeV emission into three pointlike sources (J1925+1729P, J1930+1851P, and J1932+1916P) and two extended sources (J1929+1732E and J1930+1826E), and improves significantly on the description based on the 4FGL-DR4 catalog.
Source J1932+1916P is identified as the known gamma-ray pulsar PSR J1932+1916, while J1925+1729P may be a new gamma-ray pulsar candidate distinct from the known gamma-ray pulsar PSR J1925+1720. This warrants future investigation and a search for pulsations.
Source J1930+1851P coincides with the TeV source PWN/SNR G54.1+0.3 and its GeV$-$TeV spectrum is consistent with both leptonic and hadronic interpretations, although a leptonic origin in relation to the known pulsar wind nebula (PWN) is more likely.
The GeV$-$TeV spectrum of J1929+1732E is consistent with a hybrid lepto-hadronic scenario in which the TeV emission traces the PWN powered by the pulsar PSR J1928+1746, while the GeV emission may result from interactions between particles escaped from the parent SNR and illuminating the gas environment.
Similarly, J1930+1826E is likely connected to PWN/SNR G54.1+0.3 under a hadronic scenario involving escaped particles in their early propagation stage.
Owing to spectral and/or morphological mismatches, the connection of these five GeV sources to the three LHAASO sources is not clear. This warrants deeper observations with HAWC and LHAASO, and a dedicated study of the modeling of the Galactic diffuse emission. Future CTAO observations with higher angular resolution are expected to deliver crucial information for the study of this region.

\end{abstract}


\section{Introduction} \label{intro}
 
The origin and acceleration of cosmic rays (CRs) across GeV to TeV energies, and up to the PeV regime for some Galactic sources, remain a central and unresolved problem in high-energy astrophysics \citep{Blasi2013A&ARv,Gabici2019IJMPD}. Supernova remnants (SNRs) and pulsar wind nebulae (PWNe) are considered to be the most promising candidates for accelerating CRs up to TeV–PeV energies via a variety of mechanisms such as diffusive shock acceleration, magnetic reconnection, or turbulent reacceleration. The nonthermal particles produce gamma-ray emission through both hadronic and leptonic processes. The hadronic channel involves proton and nuclei inelastic collisions and meson decay, while the leptonic channel includes electron and positron radiation from synchrotron emission and inverse Compton (IC) scattering.
Over the past decades, gamma-ray surveys from space-borne telescopes like \emph{Fermi}-LAT \citep{{Atwood+etal+2009ApJ}}, AGILE \citep{Tavani2009} and ground-based gamma-ray detectors like H.E.S.S. \citep{Hinton2004}, HAWC \citep{Abeysekara2013}, VERITAS \citep{Weekes2002}, and LHAASO \citep{Cao2019} have significantly expanded the known population of Galactic sources, many of which are spatially extended with complex morphologies \citep{Lande2012,Ackermann2017,HESS_Collaboration2018A1, Albert2020ApJ, Acharyya2023, Cao+etal+2024ApJS,Abdollahi2024}.

The Galactic plane appears particularly rich in TeV–PeV gamma-ray sources and serves as a key region for probing extreme particle acceleration and propagation. However, it also presents significant observational and interpretative challenges. The high source density often leads to confusion, especially for spatially extended sources. Moreover, the diffuse gamma-ray background remains imperfectly known due to uncertainties in the distribution of interstellar gas, magnetic field structures governing particle transport, and the possible contribution of unresolved sources \citep{Ackermann2012ApJ}. 
Disentangling these complex components requires high-resolution spatial and spectral analyses to identify the dominant emission mechanisms and to constrain the underlying particle transport conditions.

The region near Galactic longitude $\sim$53°, encompassing 1LHAASO J1928+1813u, 1LHAASO J1928+1746u, and 1LHAASO J1929+1846u \citep{Cao+etal+2024ApJS}, provides an interesting testbed for addressing these challenges. This area hosts multiple TeV gamma-ray sources detected by H.E.S.S., VERITAS, HAWC, and LHAASO, as well as GeV counterparts identified with \textit{Fermi}-LAT. Notably, it contains a number of energetic pulsars such as PSR J1925+1720, PSR J1928+1746, PSR J1930+1852, and PSR J1932+1916, along with young and powerful SNRs such as G53.4+0.0 and G54.1+0.3. The latter is considered a PeVatron (PeV CR accelerator) candidate based on its hard gamma-ray spectrum and broadband emission properties \citep{Shi2025ApJ}, which renders the region even more remarkable. All these objects reside in a complex area with overlapping emission components and a highly structured gas environment.

Clarifying the GeV–TeV connection of the different sources will help constrain the history of acceleration, escape, and propagation of CRs in these objects. In particular, the presence of several energetic pulsars in positional coincidence with the three 1LHAASO sources makes the region a compelling site for investigating the formation of diffusion suppression zones and pulsar halos \citep{Amato2024NCimR,Martin2025}, which could ultimately provide insights into the role of pulsars in contributing to the Galactic positron excess \citep{Profumo2018PhRvD, Tang2019MNRAS, Martin2022A&A}.

In this paper, we performed a comprehensive analysis of the three 1LHAASO sources region using 17.3 yr of \emph{Fermi}-LAT data. Our analysis aims to clarify the origin of the GeV gamma-ray emission and its relation to the known TeV–PeV sources, characterize the spatial and spectral properties of the emission components, and evaluate the roles of SNRs, pulsars, and surrounding interstellar medium (ISM) structures in shaping the observed gamma-ray landscape. The paper is organized as follows. Section \ref{data} describes the LAT dataset and analysis methods, and Section \ref{result} presents the corresponding results. Modeling of the gamma-ray emission and interpretation are provided in Section \ref{modeling}, while the newly identified pulsar candidate J1925+1729P is discussed in Section \ref{newPSR}. Finally, we summarized the results and concluded in Section \ref{conclusion}.

\section{Dataset AND ANALYSIS METHODS} \label{data}

We analyzed 17.3 yr of \emph{Fermi}-LAT Pass 8 data \citep{Atwood+etal+2013, Bruel+etal+2018} with reconstructed energies in the 0.1$-$500 GeV range, collected between the beginning of the mission, 2008 August 4 (MJD 54682) and 2025 December 6 (MJD 61015), corresponding to Mission Elapsed Time 239557417-786672005. SOURCE class events with the P8R3\_SOURCE\_V3 instrument response functions (IRFs) were selected and filtered with the standard event filter $\rm {(DATA\_QUAL > 0)\&\&(LAT\_CONFIG==1)}$ to get good quality data. To minimize contamination from low-energy Earth-limb emission, we varied the zenith angle cut depending on the energy range and the point-spread function (PSF) event type following the 4FGL catalog\footnote{\url{https://fermi.gsfc.nasa.gov/ssc/data/access/lat/8yr_catalog/}} \citep{4FGL}. 
We kept PSF2 and PSF3 events with zenith angles $< 90^{\circ}$ for the 0.1$–$0.3 GeV range, PSF1, PSF2 and PSF3 events with zenith angles $< 100^{\circ}$ for the 0.3$–$1 GeV range, and all events with zenith angles  $< 105^{\circ}$ for energies above 1 GeV. We also excluded time intervals affected by solar flares or gamma-ray bursts following the 4FGL-DR3 and 4FGL-DR4 catalog \citep{4FGL-DR3, 4FGL-DR4}.

With the specific zenith angle cuts for different PSF events and energy ranges, we performed summed likelihood analysis within a $14^{\circ}\times 14^{\circ}$ squared region of interest (ROI) centered on the source 1LHAASO J1929+1846u (R.A.=$292\fdg34$, decl.=$18\fdg77$), which is the position determined by the WCDA detector of LHAASO \citep{Cao+etal+2024ApJS}. 
In particular, we incorporated weights in the summed maximum likelihood analysis to mitigate systematic uncertainties arising from our imperfect knowledge of the Galactic diffuse emission following the 4FGL catalog \citep[Appendix B in][]{4FGL}. This approach yields smaller test statistics (TS; see definition later) values and larger parameter errors, providing a direct—albeit approximate—reflection of the systematic uncertainty level.
Data were binned in energy using ten logarithmic energy bins per decade, and in position with a pixel size of $0\fdg05\times0\fdg05$. 
The starting source model includes all \emph{Fermi}-LAT 4FGL-DR4 sources within a $20^{\circ}\times20^{\circ}$ squared region around the ROI center. Interstellar diffuse emissions from the Milky Way and the isotropic extragalactic and residual component were taken into account using the ``gll\_iem\_V07'' and ``iso\_P8R3\_SOURCE\_V3\_v1.txt'' models\footnote{\url{https://fermi.gsfc.nasa.gov/ssc/data/access/lat/BackgroundModels.html}}, respectively. 
The energy dispersion was applied to all sources in the model (excluding the isotropic component). 
 
The detection significance of a source is characterized by the TS, which is defined as TS$=2(\log \mathcal{L}_{1}-\log \mathcal{L}_{0})$, where $\log \mathcal{L}_{1}$ and $\log \mathcal{L}_{0}$ are the logarithms of the maximum likelihood of the complete source model and of the background model (i.e., the model without target source included), respectively \citep{Mattox+etal+1996ApJ}. Similarly, the significance of source extension is quantified by TS$_{\rm {ext}}=2(\log \mathcal{L}_{\rm e}-\log \mathcal{L}_{\rm p})$ , where $\mathcal{L}_{\rm e}$ is the logarithm of the maximum likelihood of the fit with an extended spatial model, and $\log \mathcal{L}_{\rm p}$ is that of the pointlike source model. A source is considered significantly extended in this work if TS$_{\rm ext}>16$. 

Throughout this work, we employed three spectral models for our targets: the simple power-law (PL) model (Eq.~\ref{eq:PL}), 

\begin{equation}
  \label{eq:PL}
  \frac{dN}{dE}~=~N_0 \left(\frac{E}{E_0} \right) ^{-\Gamma_0},
\end{equation}
the log-parabola (LP) model (Eq.~\ref{eq:LP}), 

\begin{equation}
  \label{eq:LP}
  \frac{dN}{dE}~=~N_0 \left(\frac{E}{E_0} \right) ^{-(\alpha+\beta \ln(E/E_0))},
\end{equation}
and the PLSuperExpCutoff4 (PLEC4) model (Eq.~\ref{eq:PLEC}), which incorporates an exponential cutoff at high energies to characterize pulsar emission following The Third \textit{Fermi}-LAT Catalog of Gamma-ray Pulsars \citep[3PC;][]{3PC}.

\begin{equation}
  \label{eq:PLEC}
  \frac{dN}{dE}~=~N_0 \left(\frac{E}{E_0} \right) ^{-\Gamma+d/b} {\rm exp} \left[\frac{d}{b^{2}}\left(1-\left(\frac{E}{E_0}\right)^{b}\right)\right],
\end{equation} 
Here,  $N_0$ represents the normalization, $\Gamma_0$ and $\alpha$ denote the spectral indices in the PL and LP models, respectively, and $\beta$ is the curvature parameter in the LP model.
 In the PLEC4 model, $\Gamma$ refers to the local spectral index at the reference energy $E_0$, $d$ represents the local curvature at $E_0$, and $b$ is the exponential cutoff index, usually set to 2/3.
The curvature of the spectra is quantified using TS$_{\rm curv} =2(\log \mathcal{L}_{\rm curv}-\log \mathcal{L}_{\rm PL})$, where $\mathcal{L}_{\rm curv}$ is the maximum likelihood of the curved model (PLEC4 or LP). In this work, the spectrum is regarded as significantly curved if TS$_{\rm curv}>9$.

%
The analysis used the Fermitools\footnote{\url{https://fermi.gsfc.nasa.gov/ssc/data/analysis/software/}} (v2.2.0) software and the $\mathtt{fermipy}$ package\footnote{\url{https://fermipy.readthedocs.io/en/latest/index.html}} \citep[v1.3.1;][]{Wood+etal+2017}.

\section{DATA ANALYSIS RESULTS} \label{result}

\subsection{Spatial Analysis} \label{morphology}

\begin{figure*}[htbp]
    \centering
        \includegraphics[width=0.48\linewidth]{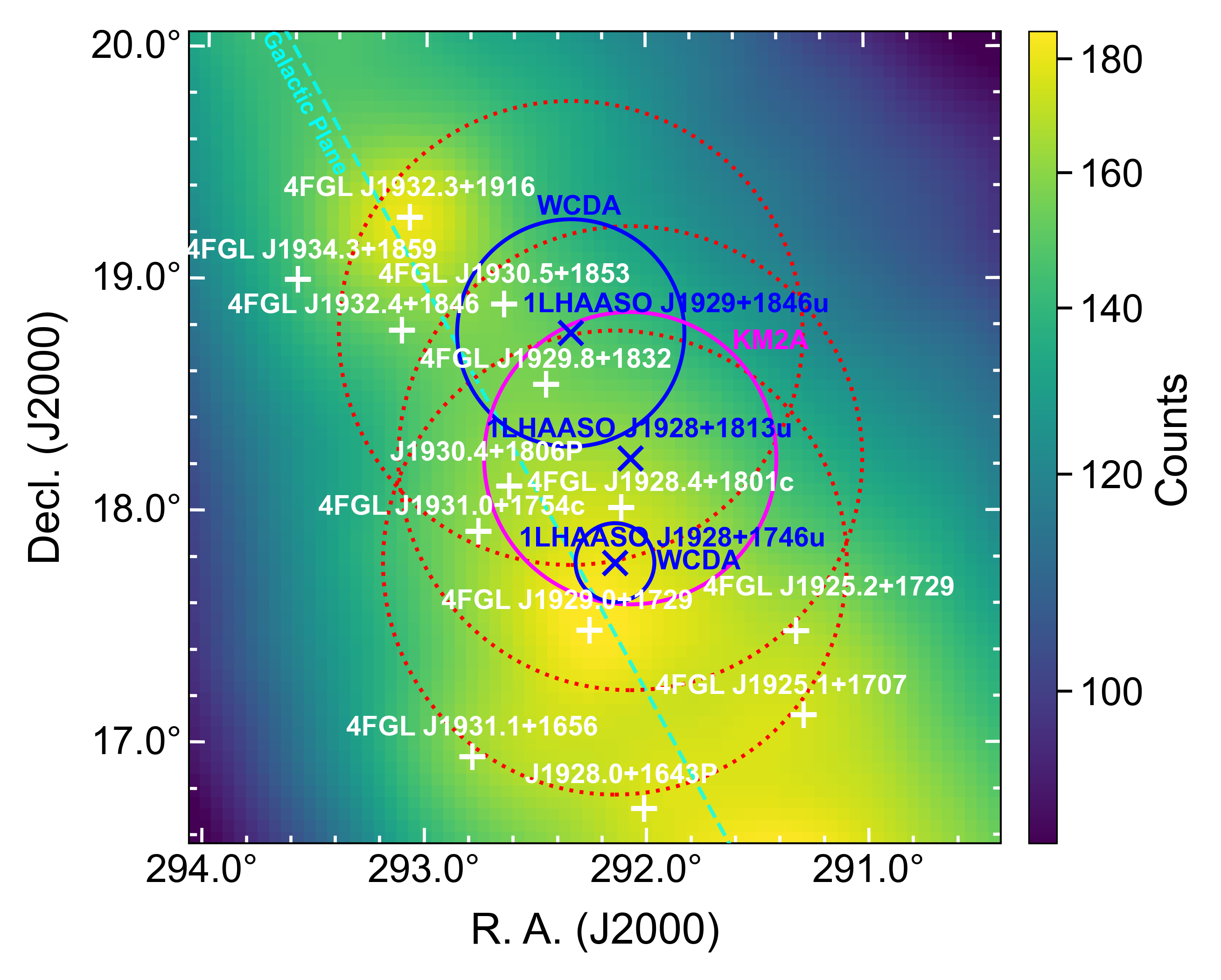}
        \includegraphics[width=0.48\linewidth]{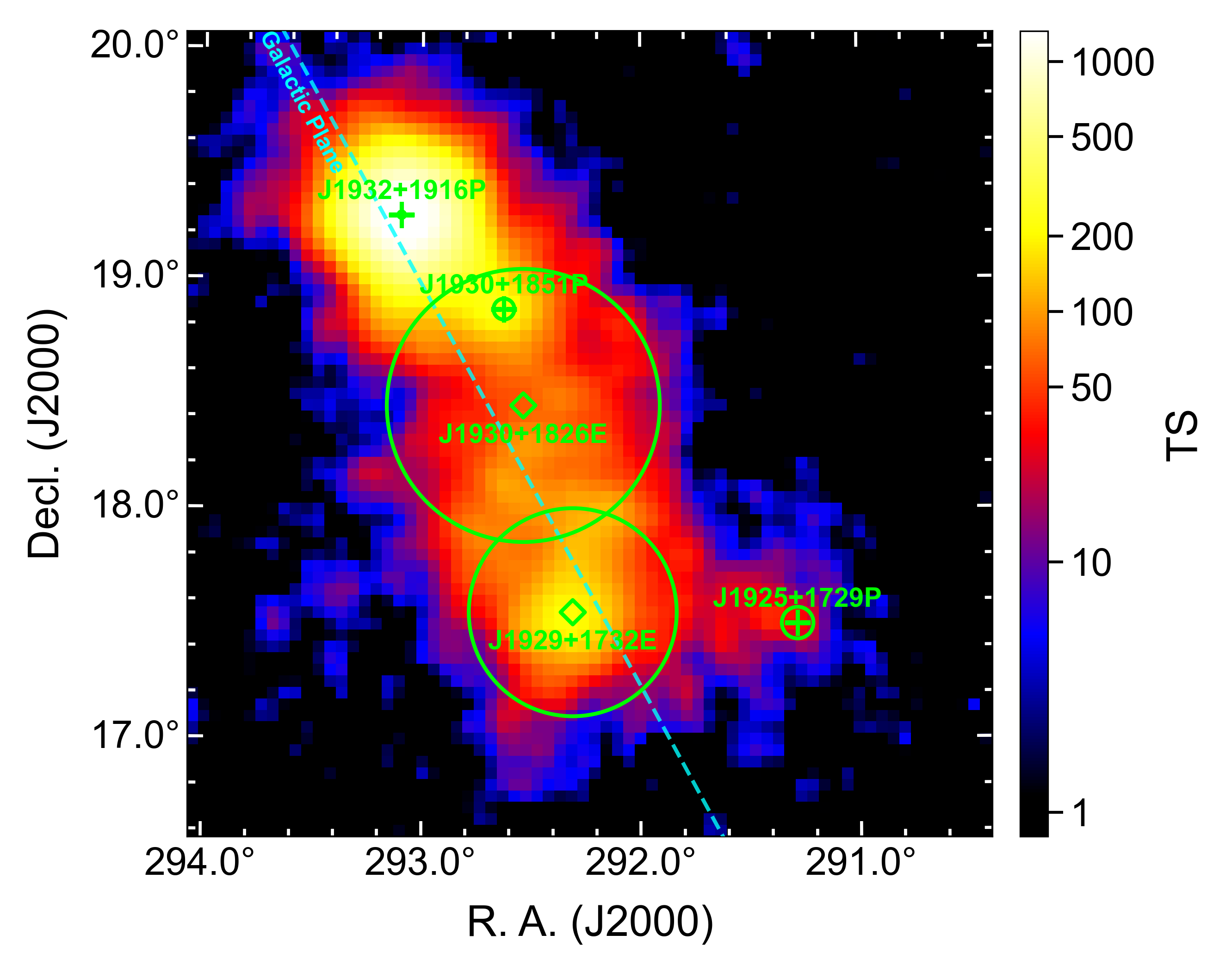} 
    \centering
        \includegraphics[width=0.48\linewidth]{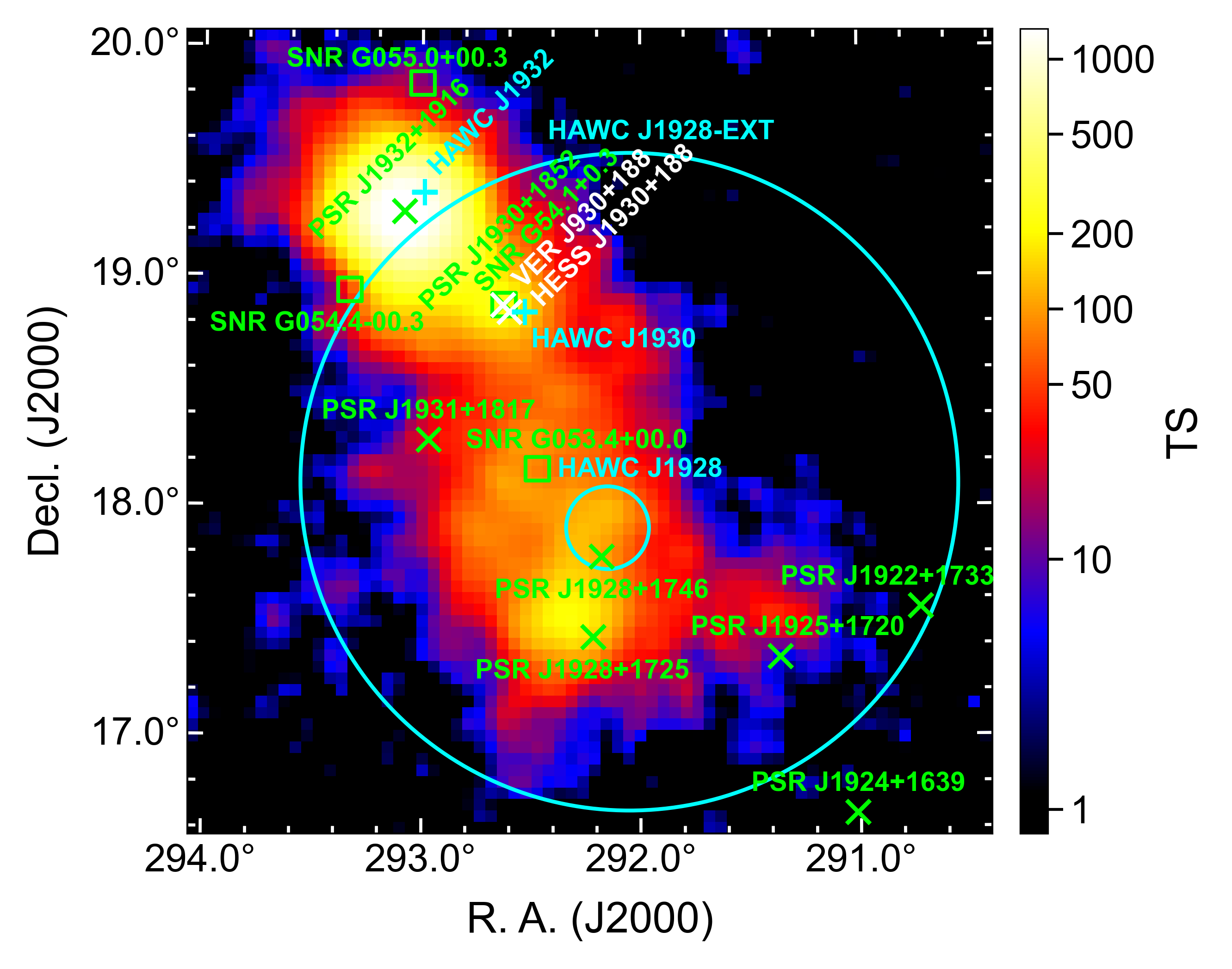}
        \includegraphics[width=0.48\linewidth]{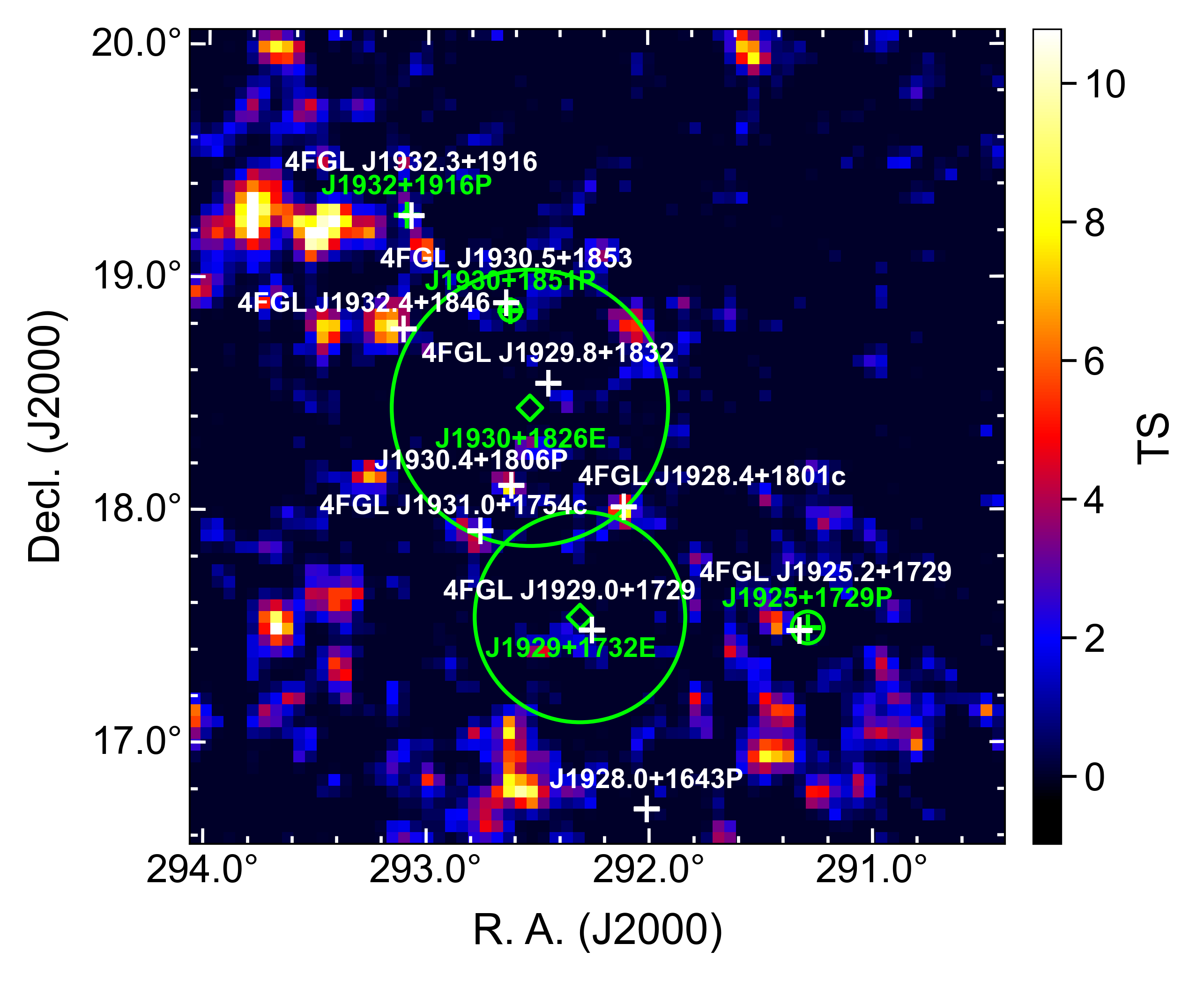}
        \caption{
        Top left: log-scaled $3\fdg5\times 3\fdg5$ smoothed counts map covering the target region in the 1$-$500 GeV energy range, with a spatial bin size of $0\fdg05$, and smoothed using a  $0\fdg15$ Gaussian kernel. Overlaid are the 4FGL-DR4 sources and two new sources found in the baseline model (white plus symbol), along with the three 1LHAASO source positions (blue cross) and the corresponding 39\% containment radii measured by the LHAASO WCDA (blue circle) and KM2A (magenta circle), as well as the $1^{\circ}$ regions around them (red dotted circle). The cyan dashed line denotes the Galactic plane.
        Top right: log-scaled TS excess map of the target region based on the initial model, showing the best-fit 3P+2G model components. Overlaid are the localizations of pointlike sources and the corresponding 95\% positional uncertainties (green plus symbol and circle), and the extended source positions and best-fit 68\% containment radii (green diamond and circle).
        Bottom left: similar to the top-right panel, but overlaid with the pulsar radio positions (green cross), SNR positions (green box), TeV associations with PSR J1930+1852 (white cross), and the latest HAWC analysis results \citep{Albert2023ApJ} of two pointlike sources HAWC J1930 and HAWC J1932 (cyan plus symbol) and the 39\% containment radii of two extended sources, HAWC J1928 and HAWC J1928-EXT (cyan circle).
        Bottom right: residual TS map after subtracting the best-fit 3P+2G model, overlaid with the model components, the 4FGL-DR4 sources, and two new sources found in the baseline model.
        }
   \label{fig:tsmap1}
\end{figure*}

The gamma-ray spatial analysis was performed by modeling the ROI in the 1$-$500 GeV energy range, which offers better angular resolution and reduced background contamination. We focus in particular on the region within $1^{\circ}$ of the three 1LHAASO sources (J1929+1846u, J1928+1813u, and J1928+1746u), which we denote as the target region. It contains eight 4FGL-DR4 sources in total, and we tried to provide an alternative and improved description of the region.
We first optimized the ROI (\emph{optimize} method in $\mathtt{fermipy}$, which iteratively fits the normalization of all sources and updates the spectral shape parameters for significant sources to approach the global likelihood maximum), then deleted weak sources (TS$<$9) and searched for new sources (TS$>$16). This resulted in one new point source in the target region (J1930.4+1806P) and 23 outside of it. Next, we performed a weighted sum likelihood fit of the ROI. 
During the fit, the spectral parameters of the sources within $5^{\circ}$ of the ROI center, and the normalizations of sources with TS$>$50 within $5^{\circ}–7^{\circ}$ of the ROI center were allowed to vary, as were the Galactic diffuse and isotropic components. 
The parameters of all other sources remained fixed to their 4FGL-DR4 catalog values. The resulting fit was adopted as the \textit{baseline model} (denoted 9P). The top left panel of Figure \ref{fig:tsmap1} shows a smoothed counts map in a $3\fdg5\times 3\fdg5$ region, overlaid with the 4FGL-DR4 sources and two new point sources (J1930.4+1806P and J1928.0+1643P) found in the above step.

Starting from the \textit{baseline model}, we first deleted the nine sources in the target region and refit the ROI. We denote this resulting model as the \textit{initial model}. Then, based on the \textit{initial model}, we systematically tested alternative models to characterize the gamma-ray emission from the target region. 
When comparing nested models, the preferred model is the one that maximizes the likelihood. 
When comparing nonnested models, we use Akaike's information criterion \citep[AIC;][]{Akaike+etal+1974}, defined as AIC $=2k - 2\log \mathcal{L}$, where $k$ represents the number of degrees of freedom in the fitted model and $\mathcal{L}$ denotes the maximum likelihood of the fit. 
The best model is the one that minimizes the AIC. 

\begin{deluxetable*}{clcccc}[htbp]
\tabletypesize{\small} 
\tablewidth{\textwidth} 
\tablecaption{\textit{Fermi}-LAT Spatial Analysis Results in the 1$-$500 GeV Energy Range
\label{tab:spatialtest}}
\tablehead{
\colhead{Model name} & \colhead{Source name} & \colhead{-$\log \mathcal{L}$} & \colhead{$k$ (d.o.f)} & \colhead{AIC} & \colhead{$\Delta {\rm AIC}$} 
} 
\startdata
& 4FGL J1932.3+1916 (P+PLEC4) &\nodata & \nodata & \nodata & \nodata \\ 
& 4FGL J1930.5+1853 (P+PL) &\nodata & \nodata & \nodata & \nodata  \\ 
& 4FGL J1932.4+1846 (P+PL) &\nodata & \nodata & \nodata & \nodata  \\ 
9P (baseline model) & 4FGL J1929.8+1832 (P+PL) & 311564.0 & 23 & 623173.9 & 0  \\
& 4FGL J1928.4+1801c (P+LP) &\nodata & \nodata & \nodata & \nodata  \\ 
& 4FGL J1931.0+1754c (P+LP) &\nodata & \nodata & \nodata & \nodata   \\ 
& 4FGL J1929.0+1729 (P+LP) &\nodata & \nodata & \nodata & \nodata  \\ 
& 4FGL J1925.2+1729 (P+PLEC4) &\nodata & \nodata & \nodata & \nodata  \\ 
& J1930.4+1806P (P+PL) &\nodata & \nodata & \nodata & \nodata  \\ 
\hline 
& J1925+1729P (P+PL) &\nodata & \nodata & \nodata & \nodata  \\ 
& J1930+1851P (P+PL) &\nodata & \nodata & \nodata & \nodata   \\ 
3P+2G (best-fit model) & J1932+1916P (P+PLEC4) &\nodata & \nodata & \nodata & \nodata  \\ 
& J1929+1732E (G+PLEC4) & 311523.8 & 24 & 623095.7 & 78.3  \\
& J1930+1826E (G+PL) &\nodata & \nodata & \nodata & \nodata  \\  
\enddata
\tablecomments{
P denotes a pointlike source, while G represents an extended source modeled with a 2D radial Gaussian. 
} 
\end{deluxetable*}

The general process of the refined analysis starting from the \textit{initial model} is summarized as follows (an alternative approach was also investigated; see later in this section):
\begin{enumerate}
\item  Start with a point source at the position of three gamma-ray pulsars, PSR J1925+1720, PSR J1928+1746, and PSR J1932+1916, and one radio pulsar PSR J1930+1852. Then, relocalize the sources (\emph{localize} method in $\mathtt{fermipy}$) and test for their extension (\emph{extension} method in $\mathtt{fermipy}$).
\item Add a new pointlike source with a PL spectrum at the position of the brightest gamma-ray excess (TS $>$ 16) in the TS map following from step 1. Then relocalize and test for its extension. Iterate this step until no additional sources (pointlike or extended) can be found with TS $>$ 16, resulting in the temporary best-fit model.
\item Test spectral curvature (\emph{curvature} method in $\mathtt{fermipy}$) for each source in the temporary best-fit model, change to PLEC4 or LP if necessary.
\item Refit the spatial and spectral parameters of all the sources in the temporary best-fit model simultaneously to get the final best-fit model. 
\end{enumerate}

\begin{deluxetable*}{ccccccccc}[htbp]
\tabletypesize{\small} 
\tablewidth{\textwidth} 
\tablecaption{Properties of the Best-fit 3P+2G Spatial Model in the 1$-$500 GeV Energy Range 
\label{tab:spatialfit}}
\tablehead{
\colhead{Source Name} & \colhead{Model} & \colhead{TS} & \colhead{TS$_{\rm ext}$} & \colhead{TS$_{\rm curv}$} & \colhead{R. A.} & \colhead{decl.} & \colhead{$p_{95}$} & \colhead{$r_{68}$} \\
\colhead{} & \colhead{(Spatial-Spectral)}& & & & \colhead{(deg)} & \colhead{(deg)} & \colhead{(deg)} & \colhead{(deg)}
}
\startdata 
J1925+1729P & P+PL & 47.3 & \nodata & 2.5 & 291.28 $\pm$ 0.03 & 17.50 $\pm$ 0.03 & 0.07 & \nodata \\ 
J1930+1851P & P+PL & 37.8 & \nodata & 0.4 & 292.62 $\pm$ 0.02 & 18.86 $\pm$ 0.02 & 0.05 & \nodata \\ 
J1932+1916P & P+PLEC4 & 1655.9 & \nodata & 67.8 & 293.10 $\pm$ 0.01 & 19.27 $\pm$ 0.01 & 0.01 & \nodata \\ 
J1929+1732E & G+PLEC4 & 377.0 & 71.8 & 28.0 & 292.31 $\pm$ 0.02 & 17.54 $\pm$ 0.03 & 0.06 & 0.45 $\pm$ 0.04  \\
J1930+1826E & G+PL & 85.8 & 58.8 & 0.0 & 292.54 $\pm$ 0.06 & 18.44 $\pm$ 0.06 & 0.15 & 0.59 $\pm$ 0.06 \\
\enddata
\tablecomments{
The model designation is the same as Table \ref{tab:spatialtest}. $r_{68}$ denotes the 68\% containment radius for extended sources, while $p_{95}$ corresponds to the uncertainty in the centroid position at a 95\% confidence interval.
}
\end{deluxetable*}

In step 1, the relocalized point source at the position of PSR J1932+1916 remained stable, whereas the source originally at the position of PSR J1925+1720 showed a clear spatial offset from the pulsar even after considering the systematic position uncertainties \citep[Section 3.2 in the 4FGL-DR3 catalog;][]{4FGL-DR3}. We discuss this point in Section \ref{newPSR}. The emission at the position of PSR J1930+1852 was found to be spatially extended, and its refined centroid was offset from the nominal pulsar location.
Similarly, the refined position of the source associated with the gamma‑ray pulsar PSR J1928+1746 also deviated from the pulsar’s position and displayed significant spatial extension. The pulsar PSR J1928+1746 was not significantly detected due to its very weak phase-averaged gamma-ray emission and thus was not included in the following analysis. We emphasize that its gamma-ray pulsations were found using the radio ephemeris as presented in 3PC.
In step 2, an additional point source was resolved, which was indeed spatially coincident with PSR J1930+1852.

In the end, the optimized model for the target region consists of three point sources and two extended sources (denoted as 3P+2G for simplicity), lying in the range of Galactic latitude $-0\fdg2 < b < 0\fdg8$.
%
%
Our best-fit model provides a better description than the \textit{baseline model} based on the 4FGL-DR4 catalog, suggesting its suitability for characterizing the observed emission. The AIC improvement is 78.3.
We summarize the results of the spatial analysis in Table \ref{tab:spatialtest} and the best-fit parameters for each source in Table \ref{tab:spatialfit}.

\begin{deluxetable*}{cccccccc}[!htbp]
\tabletypesize{\small} 
\tablewidth{\textwidth} 
\tablecaption{\textit{Fermi}-LAT Spectral Analysis Results in the 0.1$-$500 GeV Energy Range 
\label{tab3:Spectral fit}}
\tablehead{
\colhead{Source Name} & \colhead{Model} & \colhead{TS} & \colhead{$\Gamma_0$ ($\Gamma$, $\alpha$)} & \colhead{$d$} & \colhead{$\beta$} & \colhead{Photon Flux} & \colhead{Energy Flux} \\
\colhead{} & \colhead{(Spatial-Spectral)} & \colhead{} & \colhead{} &  \colhead{} & \colhead{} & \colhead{($\rm 10^{-8}~ph~cm^{-2}~s^{-1}$)} & \colhead{($\rm 10^{-12}~erg~cm^{-2}~s^{-1}$)} 
}
\startdata 
J1925+1729P & P+LP & 102.3 & 2.40 $\pm$ 0.14 &\nodata & 0.45 $\pm$ 0.09  &  1.92 $\pm$ 0.48 & 14.4 $\pm$ 2.1  \\ 
J1930+1851P & P+PL & 46.9 & 1.81 $\pm$ 0.14 & \nodata & \nodata  & 0.21 $\pm$ 0.12 & 5.9 $\pm$ 1.2  \\ 
J1932+1916P & P+PLEC4 & 2085.8 & 2.19 $\pm$ 0.04 & 0.59 $\pm$ 0.04 & \nodata  & 10.22 $\pm$ 0.73 & 68.1 $\pm$ 2.8  \\ 
J1929+1732E & G+LP & 532.9 & 2.22 $\pm$ 0.06 & \nodata & 0.38 $\pm$ 0.04  & 6.88 $\pm$ 0.79 & 56.4 $\pm$ 3.4  \\
J1930+1826E & G+LP & 90.5 & 2.11 $\pm$ 0.12 & \nodata & 0.25 $\pm$ 0.07  & 3.16 $\pm$ 0.70 & 26.6 $\pm$ 3.3  \\
\enddata
\tablecomments{$\Gamma_0$ and $\alpha$ are the spectral indices of the PL and LP models, respectively. In the PLEC4 model, $\Gamma$ is the local spectral index at $E_{0}$, $d$ is the local curvature at $E_{0}$, and the exponential cutoff index is set to $b=2/3$. In the LP model, $\beta$ is the curvature parameter. 
}
\end{deluxetable*}

The top-right and bottom-left panels of Figure \ref{fig:tsmap1} present the TS excess map covering the target region, overlaid with the best-fit 3P+2G model components and a selection of multiwavelength counterparts, respectively. 
For the HAWC counterparts used in our gamma-ray emission modeling (Section \ref{modeling}), we adopted the source names from Table 1 of \cite{Albert2023ApJ} for consistency, prefixing them with ``HAWC" to maintain a uniform notation throughout this work. This choice is motivated by the fact that their analysis represents the most recent and comprehensive description of the region as observed by HAWC. Although HAWC J1928 and HAWC J1930 correspond respectively to 3HWC J1928+178 and 3HWC J1930+188 in the third HAWC catalog \citep{Albert2020ApJ}, their positions and extensions are different from the catalog values. HAWC J1932, on the other hand, refers to the newly reported source HWC J1932+192 in \cite{Albert2023ApJ}.
The bottom-right panel shows the residual TS map obtained after subtracting the best-fit model. No significant residuals ($>$4$\sigma$) appear in the map, demonstrating the effective modeling of the region. 

We emphasize that we tested an alternative approach in the spatial analysis, by starting with a point source placed at the position of the brightest excess in the TS map (top-right panel in Figure \ref{fig:tsmap1}), i.e., at the position of the gamma-ray pulsar PSR J1932+1916, and proceeding following the same iterative process of adding, evaluating, and refining sources as described above. Ultimately, this alternative analysis yielded results that were very similar to those presented above.

\subsection{Spectral Analysis\label{spectra}}

Using the best-fit 3P+2G spatial model obtained in 1$-$500 GeV, as detailed in Table \ref{tab:spatialfit}, we performed a broadband spectral analysis within the ROI in the energy range of 0.1$-$500 GeV. Considering the larger number of photons in this extended energy band, we also first optimized the ROI, then deleted weak sources (TS$<$9, none in the target region) and searched again for new sources with TS$>$16, resulting in 39 new pointlike sources (none in the target region).
Then, we used the weighted sum likelihood to investigate the spectral properties of the gamma-ray emission from the five sources detected in our 3P+2G model.
During the fit, the spectral parameters of the sources within $3^{\circ}$ and those with TS$>$100 within $3^{\circ}–5^{\circ}$ from the ROI center were allowed to vary, while only the normalizations of sources with TS $>$ 50 in the $3^{\circ}–5^{\circ}$, along with the normalizations of sources with TS$>$100 within $5^{\circ}–7^{\circ}$ from the ROI center, were left free. This choice accounts for the balance between the broad \emph{Fermi}‑LAT PSF ($\sim 5\fdg3$ for the full PSF) at 100 MeV \citep{Principe2018,LAT10yrs} and the number of free parameters allowed. 

We reevaluated the curvature for each source and adjusted the spectral shape to either PLEC4 or LP as appropriate. Subsequently, we performed a series of fits using different combinations of spectral models and selected the best-fit model based on the AIC. Table \ref{tab3:Spectral fit} summarizes the best-fit spectral analysis results.
Note that for J1925+1729P and J1930+1826E, the LP model is only slightly better than PLEC4, with a $\Delta \log \mathcal{L}$ of $\sim$3 and $\sim$4, respectively. 
%

\begin{figure*}[!htbp]
    \centering
        \includegraphics[width=0.45\linewidth]{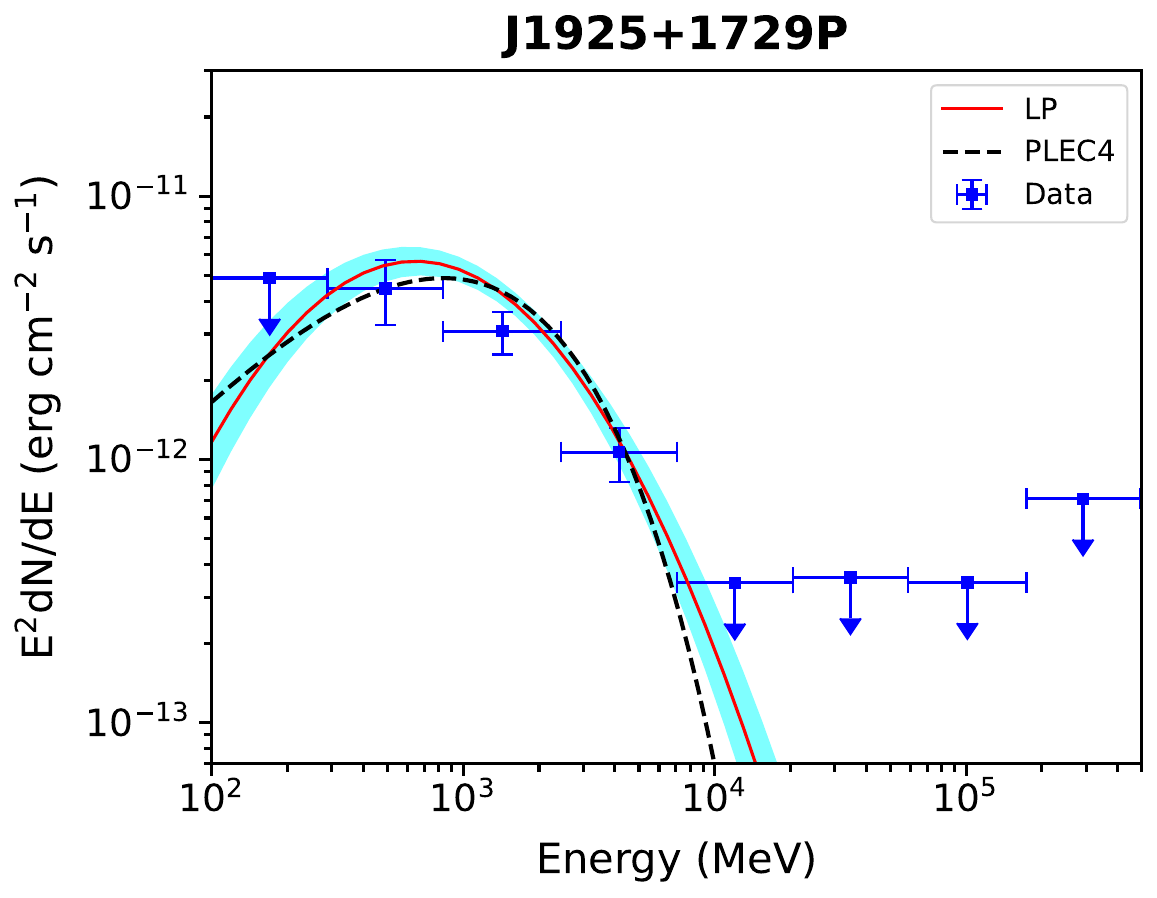}
        \includegraphics[width=0.45\linewidth]{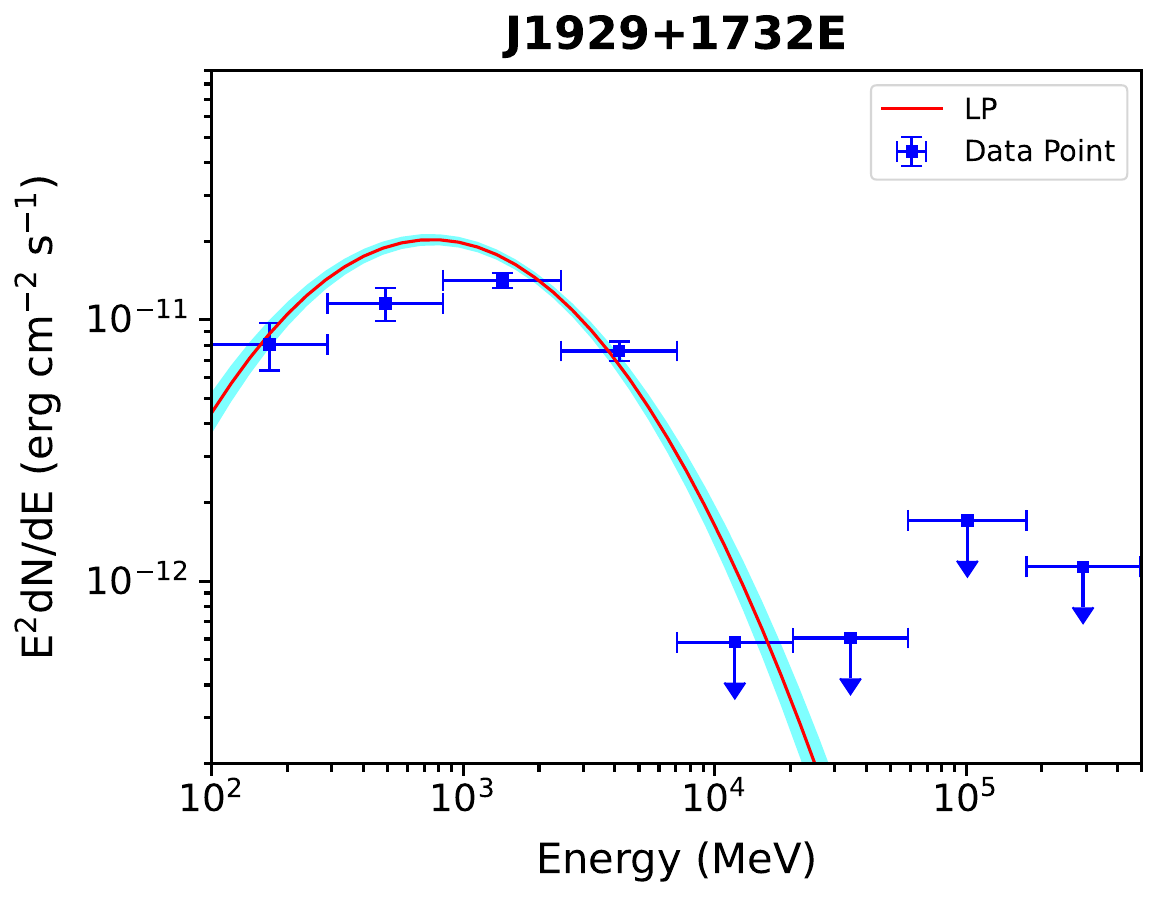}
    \centering
        \includegraphics[width=0.45\linewidth]{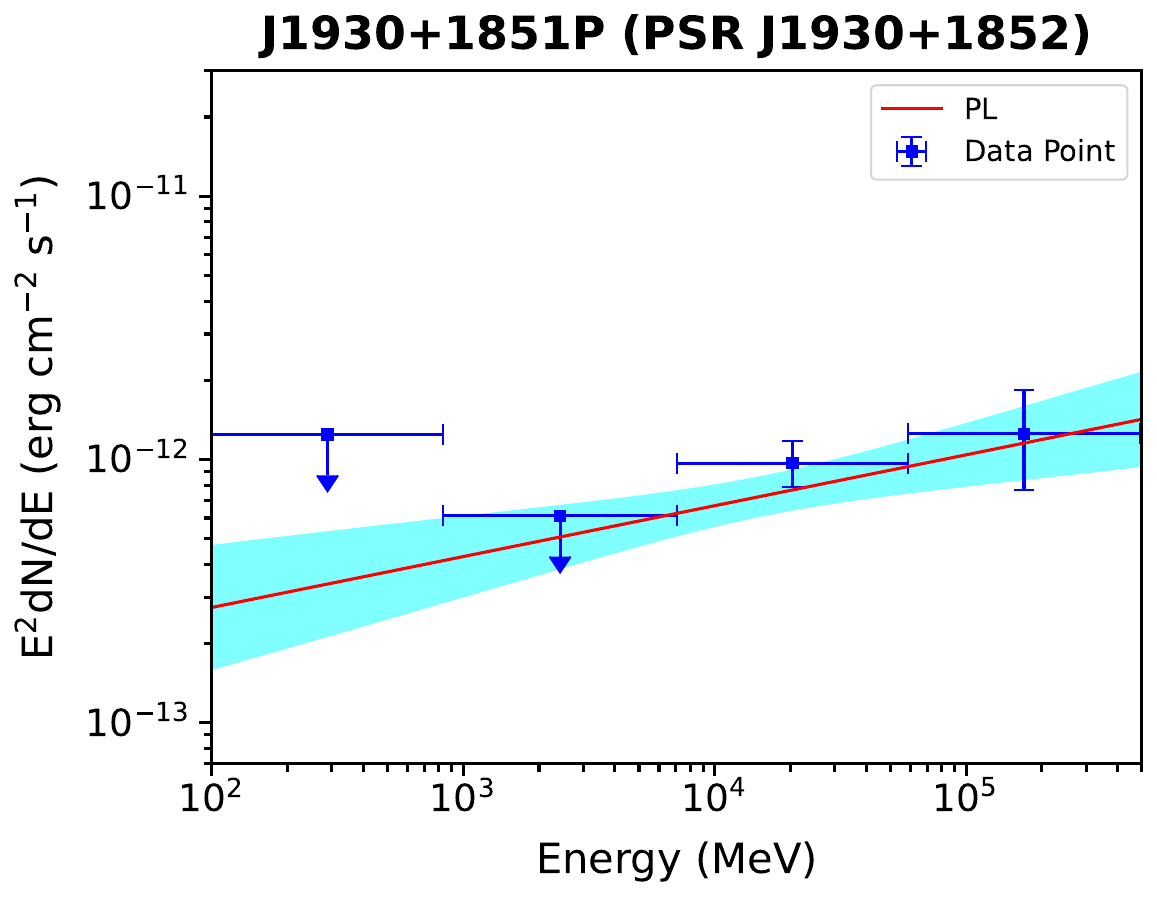 }
        \includegraphics[width=0.45\linewidth]{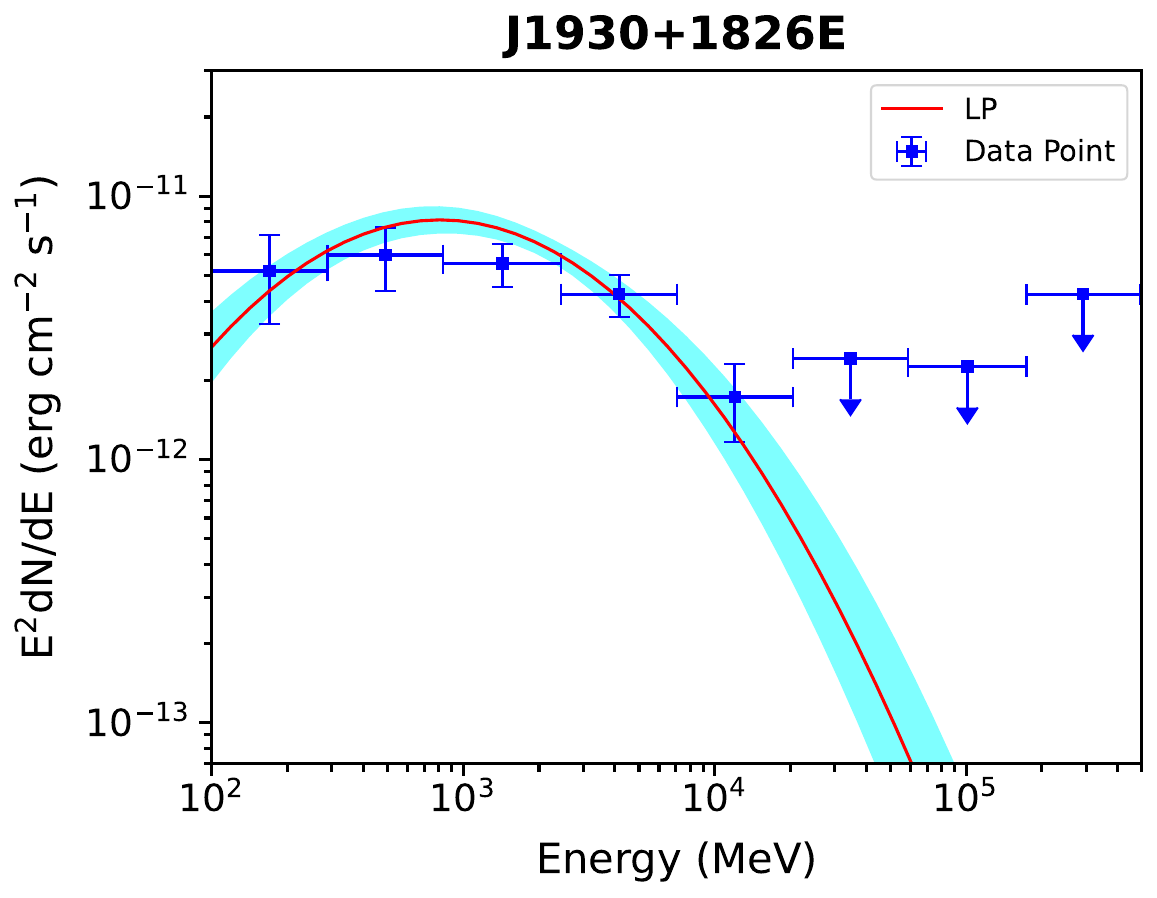}
    \centering
         \includegraphics[width=0.45\linewidth]{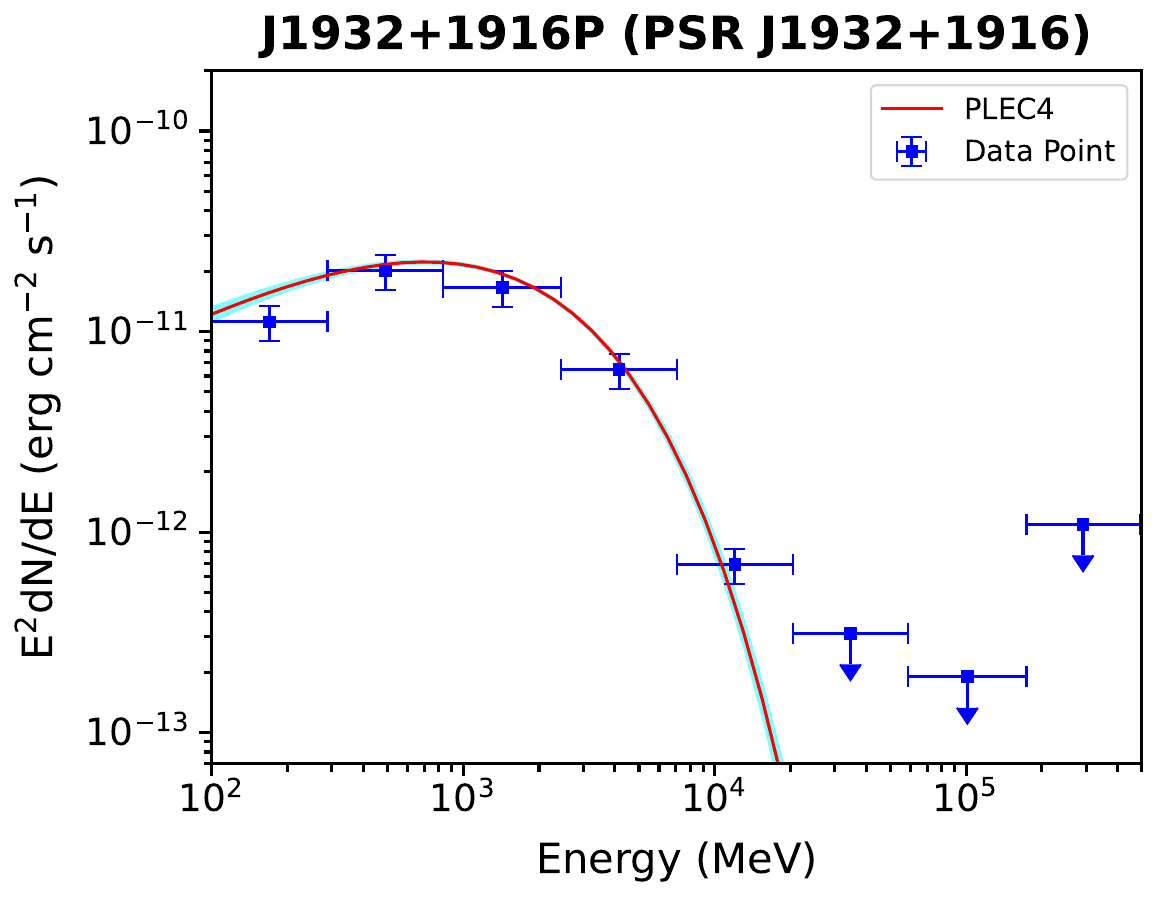}
    \caption{
    SEDs of the emission for the five detected sources in our target region. Red lines and cyan shaded regions represent the best-fit models with 1$\sigma$ statistical uncertainties, while blue data points indicate the energy flux in each energy bin, with 95\% confidence upper limits for bins with TS $<4$ shown by arrows. For J1925+1729P, the PLEC4 model (black dashed line) is also shown for comparison.
    }
    \label{fig:sed}
\end{figure*}

To obtain the spectral energy distribution (SED) of the five detected sources, we divided the data from 100 MeV to 500 GeV into eight logarithmically spaced energy bins (for J1930+1851P, four bins due to low statistics) and performed a maximum likelihood spectral analysis in each bin assuming a PL spectrum in each bin with index fixed to the local index in the broadband spectrum. The normalizations of the source of interest and background sources that were free in the broadband fit were also kept free during the fit. A 95\% confidence level upper limit was computed when the TS value was lower than 4 in a given bin. The SEDs and the best-fit broadband spectral models are shown in Figure \ref{fig:sed}.

\section{Modeling of the Gamma-Ray Emission}
\label{modeling}

To investigate the possible physical origin of the gamma-ray emission of our target sources, we modeled the \emph{Fermi}-LAT spectra obtained in this work together with HAWC spectra when available, using the \texttt{naima}\footnote{https://github.com/zblz/naima} package \citep[v0.10.2;][]{Zabalza2015ICRC}. This package uses different radiative models to compute nonthermal emission from relativistic particle populations with a given spectral distribution, and the models are fitted to the observed multiwavelength spectra via a Markov Chain Monte Carlo \citep{Foreman-Mackey2013}. 
For the HAWC spectra, mock SED data points were derived from the spectral functions and energy ranges reported in Table 1 of \citet{Albert2023ApJ}. 
Among the five sources we detected (Table \ref{tab3:Spectral fit}), J1932+1916P most likely corresponds to the gamma-ray PSR J1932+1916, whose gamma-ray emission in the LAT band is predominantly magnetospheric, while the gamma-ray emission of the TeV counterpart HAWC J1932 may come from accelerated particles in the PWN powered by the same pulsar, as investigated in \citet{Albert2023ApJ}. As discussed in Section \ref{newPSR}, source J1925+1729P may well be a new pulsar, with a GeV spectrum typical of magnetospheric emission, that lacks a HAWC counterpart. We ignore the latter two sources in our modeling exercise (because magnetospheric emission requires a more involved approach), and focus instead on J1930+1851P, J1930+1826E, and J1929+1732E.

We considered two main radiative processes for the gamma-ray emission of the three target sources: IC scattering of relativistic electrons and positrons on ambient photons (leptonic model) or neutral pion decay (PD) from CR proton and nuclei interactions (hadronic model). Although bremsstrahlung emission from electrons can also contribute, we neglected it here since its effect becomes negligible above a few GeV for typical ambient gas densities \citep[$n_{\mathrm{H}}\lesssim100~\mathrm{cm^{-3}}$;][]{Aharonian2006A&A}.
In the IC model, the far infrared to ultraviolet target photon fields were taken, in the direction and at the distance of each source, from the Galactic interstellar radiation field model of \citet{Popescu2017MNRAS}, which we complemented with the cosmic microwave background (CMB; with blackbody temperature $T = 2.72~\mathrm{K}$ and energy density $U_{\mathrm{CMB}} = 0.26~\mathrm{eV~cm^{-3}}$). 

The underlying particle populations were parameterized by different spectral forms: a simple PL, possibly including an exponential cutoff (ECPL),

\begin{equation}
  \label{eq:ECPL}
  \frac{dN}{dE}~=~N_0 \left(\frac{E}{E_0} \right) ^{-\Gamma}{\rm~exp}\left(-\frac{E}{E_c} \right)\, ,
\end{equation}
or a broken PL (BPL),
\begin{equation}
  \label{eq:BPL}
  \frac{dN}{dE} = N_0 
  \begin{cases} 
  \left( \frac{E}{E_0} \right)^{-\Gamma} & E < E_b \\[6pt]
  \left( \frac{E_b}{E_0} \right)^{\Gamma_1-\Gamma} \left( \frac{E}{E_0} \right)^{-\Gamma_1} & E > E_b \,\, ,
  \end{cases}
\end{equation}
or a BPL with an exponential cutoff (ECBPL),
\begin{equation}
  \label{eq:ECBPL}
  \frac{dN}{dE} = N_0 {\rm~exp}\left(- \frac{E}{E_c} \right)
  \begin{cases} 
  \left( \frac{E}{E_0} \right)^{-\Gamma} & E < E_b \\[6pt]
  \left( \frac{E_b}{E_0} \right)^{\Gamma_1-\Gamma} \left( \frac{E}{E_0} \right)^{-\Gamma_1} & E > E_b  \,\,,
  \end{cases}
\end{equation}
in which the cutoff energy $E_c$, the break energy $E_{b}$, and the spectral indices below and above the break, $\Gamma$ and $\Gamma_{1}$, are left free. The cutoff energy $E_c$ represents the characteristic energy at which the particle spectrum starts to decline exponentially, possibly indicating the maximum energy achievable in a given accelerator, and the reference energy $E_0$ is set to a fixed value (e.g., 1 GeV, 1 TeV, or 10 TeV) for different models.

For each source, the overall SED may be explained by a single particle population and a single IC or PD process, or by more than one particle population and a combination of emission processes. In this work, our strategy is to first test the simplest scenario, i.e., a single particle population and a single process, and then to investigate more complex models only if the simplest scenario fails to satisfactorily reproduce the data. 
This approach allows us to probe the particle properties while keeping the interpretation as model-independent as possible, given the limited knowledge of the sources and their environments. The best-fit parameters of the particle distribution are summarized in Table \ref{tab:sed modeling}, and Figure \ref{fig:naima_sed} shows the resulting photon spectra along with the GeV or GeV$-$TeV SEDs of the target sources. The large upper uncertainty in the total electron energy $W_{e}$ for J1930+1851P and J1929+1732E in the IC scenario primarily arises from poorly constrained electron spectra at low energies, as the spectral fit is predominantly sensitive to the cutoff region. For J1930+1851P, a BPL model can fit the data as well as the ECPL model, with both models predicting a strong drop in the particle population at $\sim~10 – 100$ TeV. In the case of J1930+1826E, the cutoff energy $E_{\rm c,p}$ in the PD process is poorly constrained, while a simple PL model provides a comparable fit.

\renewcommand{\arraystretch}{1.5} 
\begin{deluxetable*}{lcccccc}
\tablecaption{Best-fit Theoretical Model Parameters for Target Sources \label{tab:sed modeling}}
\tabletypesize{\small}
\tablewidth{0pt}
\tablehead{
\colhead{Best-Fit Model} & $\Gamma_{\rm e}$ & $\Gamma_{\rm p}$ & \colhead{$E_{\rm c,e}$} & \colhead{$E_{\rm c,p}$} & \colhead{$W_{e}$} & \colhead{$W_{p}$} \\
 \colhead{} & \colhead{} & \colhead{} & \colhead{(TeV)} & \colhead{(TeV)} & \colhead{(erg)} & \colhead{(erg)}
}
\startdata
\multicolumn{7}{c}{J1930+1851P + HAWC J1930} \\
\hline
ECPL+IC & $2.42^{+0.32}_{-0.42}$ & \nodata & $25^{+37}_{-12}$ & \nodata & $1.5^{+16.1}_{-1.4}\times10^{49}$ & \nodata \\
ECPL+PD & \nodata & $1.91^{+0.11}_{-0.08}$ & \nodata &  $87^{+72}_{-28}$ & \nodata & $1.0^{+0.2}_{-0.2}\times10^{49}$   \\
\hline
\multicolumn{7}{c}{J1930+1826E} \\
\hline
ECPL+PD & \nodata & $2.50^{+0.15}_{-0.30}$ & \nodata & $1^{+20}_{-1}$ & \nodata & $9.1^{+5.9}_{-4.6}\times10^{49}$  \\
\hline  
\multicolumn{7}{c}{J1929+1732E + HAWC J1928} \\
\hline                      
ECPL+PD+IC & $2.18^{+0.35}_{-0.36}$ & $1.95^{+0.41}_{-0.27}$ & $130^{+44}_{-36}$ & $0.04^{+0.13}_{-0.01}$ & $0.4^{+9.9}_{-0.3}\times10^{48}$ & $3.1^{+2.7}_{-0.9}\times10^{49}$  
\enddata
\tablecomments{$\Gamma_{\rm e}$ and $E_{\rm c,e}$ correspond to the particle spectral index and the cutoff energy of the electron component, while $\Gamma_{\rm p}$ and $E_{\rm c,p}$ denote the particle spectral index and cutoff energy of the proton component. $W_e$ and $W_p$ represent the total energy in electrons and protons calculated above 100 MeV.  }

\end{deluxetable*}

\subsection{Origin of J1930+1851P}

Pointlike source J1930+1851P is spatially coincident with PSR J1930+1852 \citep{Camilo2002} and its TeV PWN G54.1+0.3, as for the original catalog source 4FGL J1930.5+1853. 
At TeV energies, it also coincides with VER J1930+188, HESS J1930+188, HAWC J1930 \citep{Abeysekara2018ApJ,HESS_Collaboration2018A1, Albert2023ApJ}, 1LHAASO J1929+1813u, and 1LHAASO J1929+1846u \citep{Cao+etal+2024ApJS}. The ultra-high-energy (UHE; $>100$ TeV) gamma-ray emission of the latter has been proposed to originate from G54.1+0.3 \citep{Xia2023RAA,Shi2025ApJ}. Yet, since the LHAASO sources are extended in nature, we will rather consider them in relation to the extended source that we detected with the LAT in our target region (see below).
%

Multiwavelength studies show that G54.1+0.3 is a Crab-like young composite PWN/SNR system \citep{Reich1985,Velusamy1988,Lu2002} powered by PSR J1930+1852. The PWN is morphologically associated with a molecular cloud with a CO emission peak at a velocity of $\sim53~{\rm km~s^{-1}}$ \citep{Leahy2008AJ}. The latter can be translated into a distance of 4.9 kpc using the latest rotation curve model in \cite{Reid2019}\citep[see][]{Shi2025ApJ}, and we adopted this distance in our modeling.
Given that both J1930+1851P and HAWC J1930 are spatially coincident with G54.1+0.3, we assumed that they are the same source and jointly modeled the \emph{Fermi}-LAT and HAWC spectra. We found that both pure leptonic and pure hadronic models can provide decent descriptions of the observed broadband spectrum.

\begin{figure*}[htbp]
    \centering
        \includegraphics[width=0.48\linewidth]{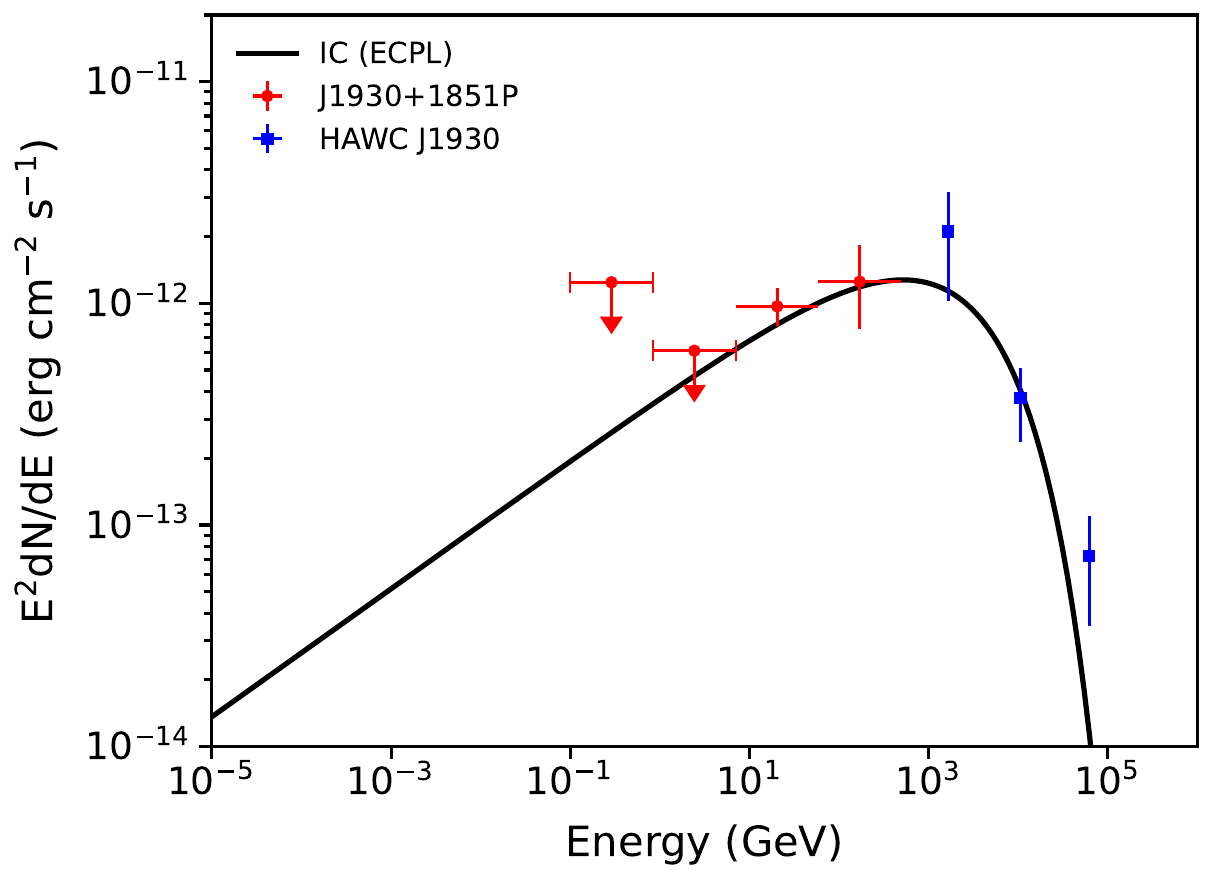}
        \includegraphics[width=0.48\linewidth]{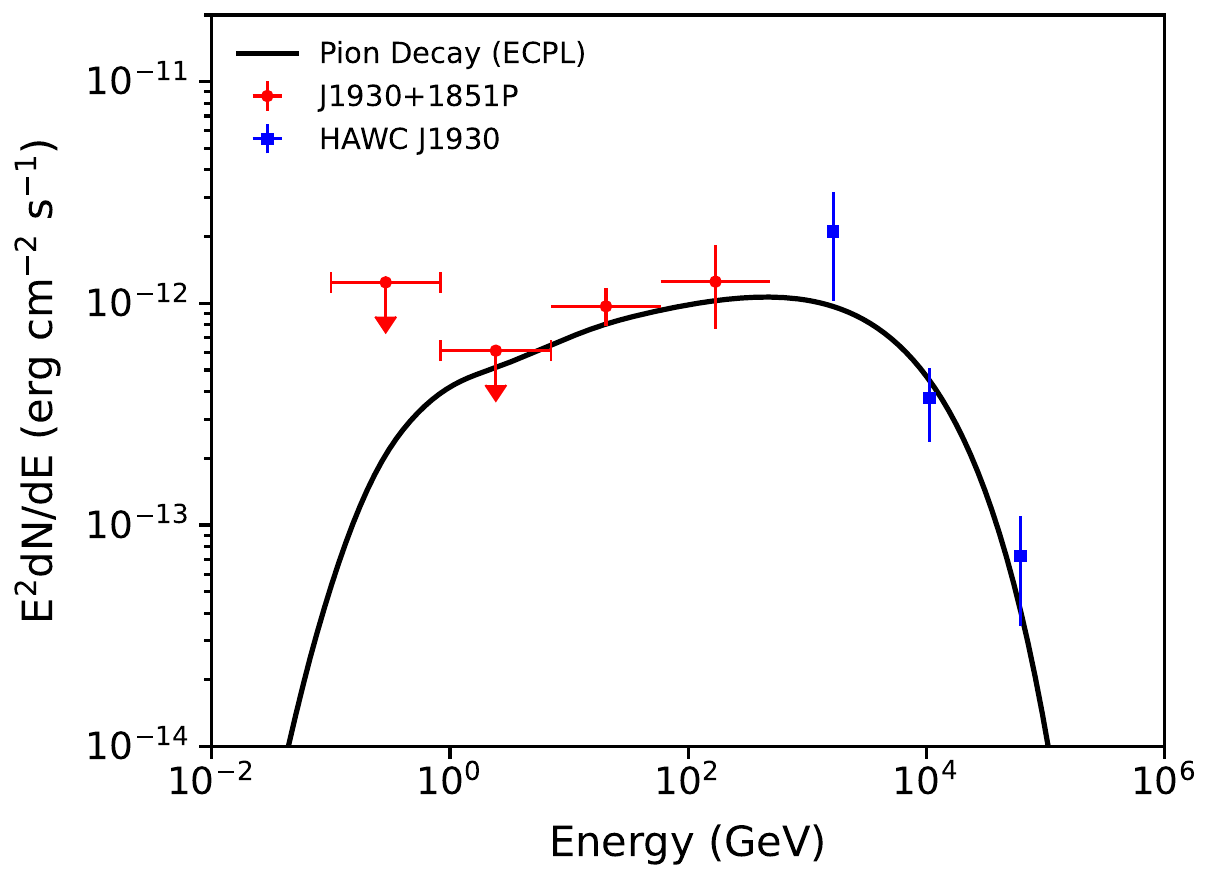}
    \centering
        \includegraphics[width=0.48\linewidth]{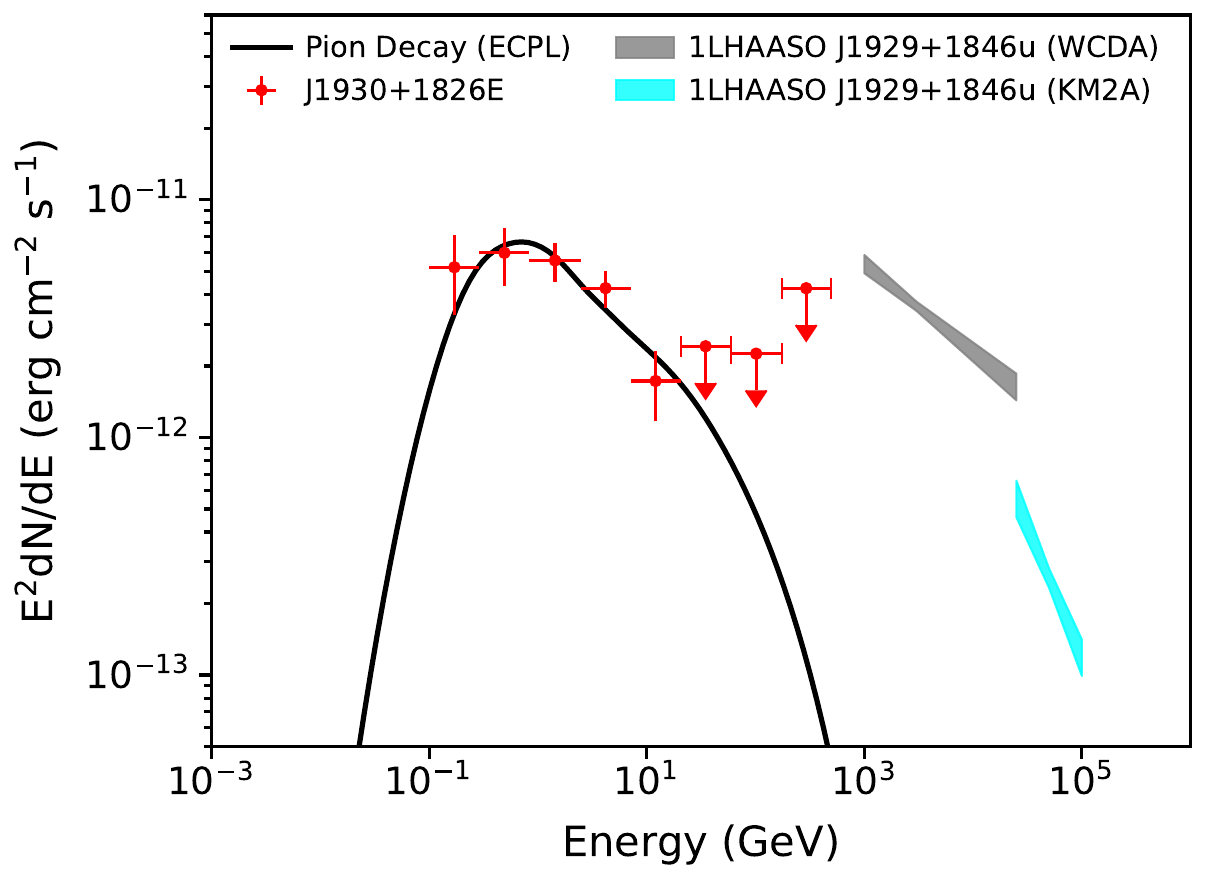}
        \includegraphics[width=0.48\linewidth]{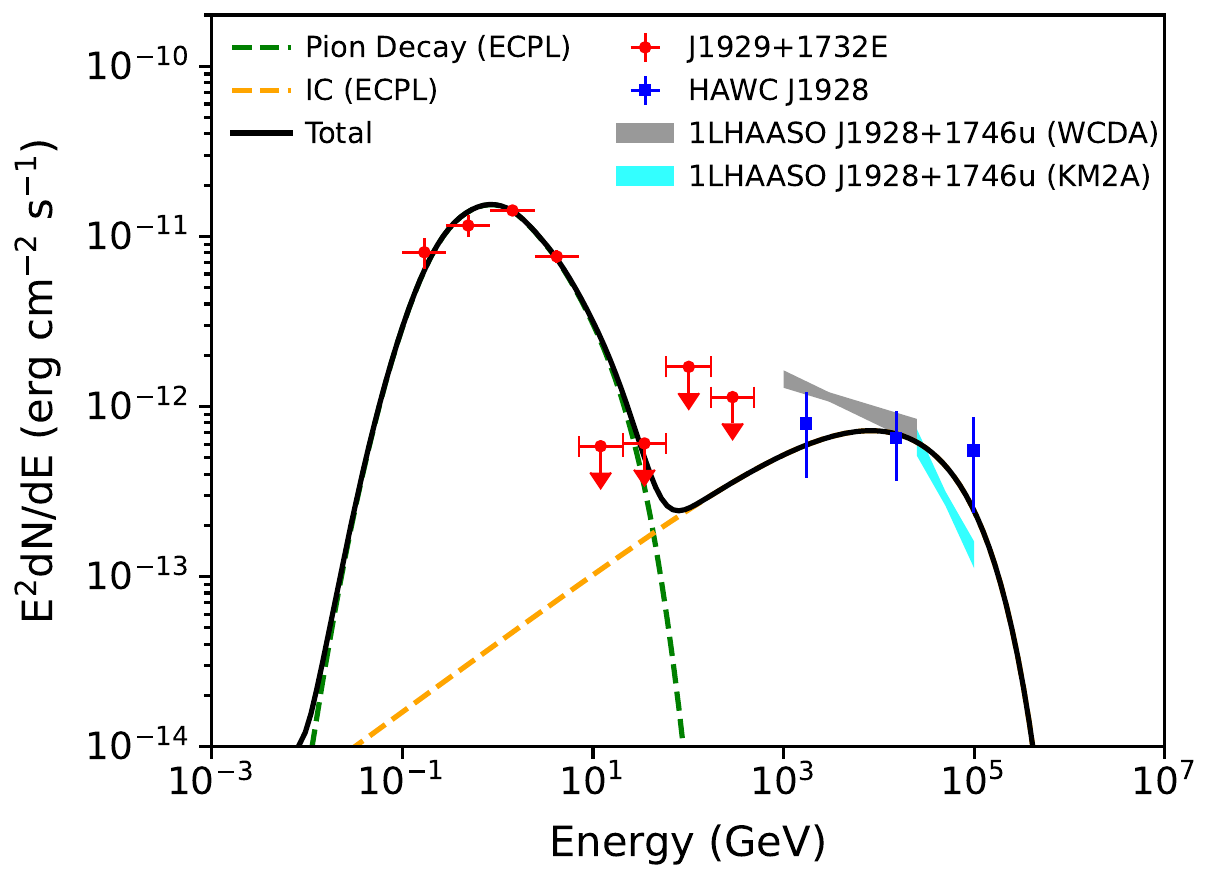}
    \caption{Broadband SED modeling of the three target sources.
    Top left: leptonic model for J1930+1851P jointly fitted with HAWC J1930, assuming an electron distribution with an ECPL spectrum.
    Top right: hadronic model for J1930+1851P jointly fitted with HAWC J1930, assuming a proton distribution with an ECPL spectrum.
    Bottom left: hadronic model for J1930+1826E, assuming a proton distribution with an ECPL spectrum.
    Bottom right: hybrid lepto-hadronic model for jointly fitted J1929+1732E and HAWC J1928. The green and orange dashed lines represent the PD and IC photon spectra derived from the same ECPL particle distribution, respectively.
    The black lines in four panels represent the total photon spectra derived from the particle distribution. The red dots represent the GeV fluxes obtained in this work, and blue squares are the HAWC fluxes of HAWC J1928 and HAWC J1930 reconstructed from the spectral functions reported in Table 1 of \citet{Albert2023ApJ}. The gray and cyan shaded bands represent the fluxes and 68\% confidence uncertainties of 1LHAASO catalog sources detected by WCDA and KM2A of LHAASO \citep{Cao+etal+2024ApJS}. 
    }
    \label{fig:naima_sed}
\end{figure*}

In the leptonic scenario, the GeV$-$TeV emission from J1930+1851P and HAWC J1930 can be well reproduced by IC emission from electrons with an ECPL distribution (see top-left panel of Figure \ref{fig:naima_sed}), supporting the physical association between the GeV and TeV sources. The best-fit model yields a total electron energy above 100 MeV of $W_{e} = 1.5\times10^{49}$ erg, which we now compare to the maximum available pulsar energy.
From the current spin-down power $\dot{E}=1.2\times10^{37}~\mathrm{erg~s^{-1}}$ \citep{Camilo2002} and characteristic age $\tau_c=2.9$ kyr of PSR J1930+1852, we can estimate the pulsar’s total rotational energy released over its lifetime. 
Under the assumption of a small initial spin-down timescale (i.e., $\tau_0 = 10–100$ yr), which maximizes the pulsar energetics, the current age of the pulsar is $t=\tau_c-\tau_0 \simeq (2.8-2.9)$ kyr and its initial spin-down luminosity $\dot{E}_0$ can be derived from
\begin{equation}
  \label{eq:cal_E_dot}
  \dot{E}(t) = \dot{E}_0 \left(1+\frac{t}{\tau_0}\right)^{-\frac{n+1}{n-1}} \,\, ,
\end{equation}
where $n$ is the braking index (taken as 3). This results in $\dot{E}_0 \simeq 10^{40}-10^{42}~\mathrm{erg~s^{-1}}$. Consequently, the maximum possible rotational energy released by PSR J1930+1852 over its lifetime could reach $E_{\mathrm{tot}}= \dot{E}_0\times\tau_0 \simeq (3.2\times10^{49}$--$3.2\times10^{50})~\mathrm{erg}$, with the larger values corresponding to the smaller $\tau_0$. 
In this most optimistic scenario, PSR J1930+1852 is therefore able to power the population of leptons that radiate today. Taken at face value, the corresponding energy-conversion efficiency is $\eta = W_{e} / E_{\mathrm{tot}} \simeq 4.7\%-47.2\%$, with the larger efficiency obtained with a larger $\tau_0$. Yet, such an efficiency should only be considered a lower limit since we did not account for the energy losses experienced by particles since they were injected millennia ago (predominantly from adiabatic losses and synchrotron emission in the nebular magnetic field).

We also considered a hadronic scenario as an alternative to the leptonic one. Recently, radio observations using MeerKAT at 1.3 GHz revealed a circular feature $\sim 14'$ in diameter surrounding PSR J1930+1852 and its PWN G54.1+0.3, and this was proposed as a candidate SNR shell associated with the pulsar \citep{Goedhart2024}. This is the first clear detection of the shell after \cite{Lang2010}. This remnant shell and the molecular cloud at a velocity of $\sim 53~{\rm kms}^{-1}$ provide a favorable target environment for the PD process. In this framework, the ambient target gas density near G54.1+0.3 was assumed to be $n_{\rm H} = 10~{\rm cm}^{-3}$ \citep{Ren2019MNRAS}.
The PD model yielded an equally good fit to the GeV$-$TeV spectra as shown in the top-right panel of Figure \ref{fig:naima_sed}. The total energy in protons above 100 MeV was estimated to be $W_{p} \simeq 1.0\times10^{49}$ erg, corresponding to an energy-conversion efficiency of $\sim1.0\%$ when compared with the canonical kinetic energy of an supernova explosion $E_{\rm SN}\simeq10^{51}$ erg \citep{Blasi2013A&ARv}. This value is consistent with the canonical $\sim5\%-10\%$ usually used in diffusive shock acceleration when considering uncertainties of order unity in the target gas density.

Our result indicates that G54.1+0.3 could act as a potential hadronic accelerator capable of producing gamma-ray emission up to multi-TeV energies. 
Our hadronic interpretation agrees with the lepton-hadron hybrid model proposed by \citet{Li2010MNRAS}, in which TeV gamma rays originated mainly from the PD process produced by interactions between protons accelerated in the PWN/SNR system and a nearby molecular cloud. 

Yet, the hard spectrum at $\lesssim1$ TeV energies is somewhat unusual in this hadronic SNR interpretation and is more consistent with the typical leptonic spectra of PWNe \citep{Torres2014JHEAp}, unless the observed emission is dominated by a higher-energy particle population resulting from energy-dependent escape and illuminating the gaseous environment of the system \citep{Magic2023A&A}. In this scenario, the suppressed GeV emission and the apparently low efficiency naturally arise because the observations probe only the escaped, high-energy component of the CR population.

\subsection{Origin of J1930+1826E}

Source J1930+1826E is extended with a 68\% containment radius of $0\fdg59 \pm 0\fdg06$ (assuming a 2D Gaussian morphology). As shown in Figure~\ref{fig:tsmap1}, J1930+1826E overlaps J1930+1851P, PSR J1930+1852, PWN/SNR G54.1+0.3, and their several TeV counterparts. It also significantly overlaps 1LHAASO J1929+1813u and 1LHAASO J1929+1846u. However, we dismiss 1LHAASO J1929+1813u because this very extended component, detected only with KM2A, did not survive in a refined analysis of this region from LHAASO data by some of us (publication in preparation). Finally, no HAWC counterpart to J1930+1826E is reported in \cite{Albert2023ApJ}, except the very extended J1928-EXT component that may be due to large-scale Galactic diffuse emission according to the authors. Owing to its extended nature and the possible continuation of its emission to tens of TeV, J1930+1826E might be powered by particles energized in PWN/SNR G54.1+0.3 that escaped the accelerator and are currently propagating away. A similar scenario may be considered in connection with another remnant, SNR G53.4+0.0.

We first considered a scenario in which J1930+1826E is physically associated with PWN/SNR G54.1+0.3 and located at the same distance of 4.9 kpc used for J1930+1851P. Under this assumption, the observed angular separation between the center of J1930+1826E and G54.1+0.3 ($\theta = 0\fdg44$) corresponds to a projected physical length of $d \simeq D \times \theta \simeq 38$ pc, where $D$ = 4.9 kpc and $\theta$ is expressed in radians.
This can be compared to the diffusion radius expressed as
$R_{\rm diff} \approx \sqrt{4 D_{\rm ISM}t}$, adopting a standard ISM diffusion coefficient at 10 GeV of $D_{\rm ISM} \approx 3 \times 10^{28}\,\rm cm^2\,s^{-1}$ \citep{Aharonian1996A&A}, and the age of G54.1+0.3 ($\tau_{c} \approx 2900$ yr). This yielded $R_{\rm diff} \approx 34$ pc, comparable to but slightly smaller than the projected separation of $\sim38$ pc, 
indicating that the physical association is plausible for particle diffusion proceeding in conditions similar to the large-scale Galactic average.

Motivated by the spatial coincidence and the presence of CO molecular clouds in this region, we investigated a hadronic origin for the GeV gamma-ray emission. Assuming a target gas density of $n_{\rm H}=10~{\rm cm^{-3}}$ as in the modeling for J1930+1851P, the PD model fit shows good consistency with the GeV gamma-ray data, and the total energy in accelerated protons above 100 MeV was found to be $W_{p}\simeq2.3\times10^{50}$ erg.
This exceeds a bit the canonical $\sim5\%-10\%$ usually used in diffusive shock acceleration. By adopting a slightly higher gas density of $n_{\rm H}=25~{\rm cm^{-3}}$, the required proton energy reduced to a reasonable value of $W_{p}\simeq 9.1\times10^{49}$ erg, making the SNR scenario energetically possible.
%
%
The fitted PL index for the proton distribution is $\simeq2.5$, in line with the soft distributions inferred for the majority of young or interacting SNRs \citep{Acero2013ApJS}, possibly suggesting that the whole CR budget was released and is now propagating out.

In addition, Figure~\ref{fig:naima_sed} shows that the GeV gamma-ray spectrum of J1930+1826E does not smoothly connect to the 1$–$25 TeV and 25$–$100 TeV spectra of 1LHAASO J1929+1846u (from WCDA and KM2A, respectively), despite the relatively similar sizes of the sources ($r_{68} = 0\fdg59$ for J1930+1826E and $r_{39} = 0\fdg49$ for 1LHAASO J1929+1846u), but keeping in mind the $0\fdg3-0\fdg4$ offset in their centroids.
Therefore, a second emission component is needed to account for the TeV emission from 1LHAASO J1929+1846u. A possible explanation for this mismatch may relate to the modeling of the Galactic diffuse emission, particularly critical in this gas-rich region. Actually, it may also be that both J1930+1826E and 1LHAASO J1929+1846u reflect imperfections in the diffuse modeling, but in different proportions.
In that respect, a refined analysis of LHAASO observations, based on more collected data and informed by our results, is underway and will be instrumental in clarifying the layout of the emission and its connection to the different objects in the region.

Last, we also examine a scenario in which J1930+1826E is associated with SNR G53.4+0.0 at a distance of 7.5 kpc \citep{Driessen2018}. 
The angular separation between J1930+1826E and G53.4+0.0 ($\theta = 0\fdg29$) corresponds to a projected physical separation of approximately 38 pc, assuming both sources are located at the same distance. 
Given that G53.4+0.0 is identified as a young SNR \citep[$\tau_{c}\sim1–5$ kyr,][]{Domcek2022}, the corresponding diffusion radius under average ISM conditions is $R_{\rm diff} \approx 20–45$ pc. In this case, diffusive transport is also consistent with the inferred separation.
However, the average ISM diffusion coefficient likely represents an upper limit, as particle diffusion around young SNRs is expected to be suppressed ($D \ll D_{\rm ISM}$) by self-generated turbulence \citep[e.g.,][]{Gabici2009,Fujita2010,Malkov2013}, which would reduce the diffusion radius and possibly exclude the observed separation out of the inferred range of $R_{\rm diff}$.

In addition, X-ray studies of G53.4+0.0 indicate a relatively low shocked gas density of $n_{\rm H}\sim0.8-0.9~\mathrm{cm^{-3}}$ \citep{Driessen2018, Domcek2022}. 
Thus, the combination of the larger distance to the source and the lower gas density would require a total proton energy of around $\simeq 6.6\times10^{51}$ erg for $n_{\rm H}\sim0.8~\mathrm{cm^{-3}}$, well exceeding the total kinetic energy of a typical supernova explosion ($\sim10^{51}$ erg).
Such an extreme energetic requirement 
would make the connection of J1930+1826E with SNR G53.4+0.0 less likely in a hadronic scenario, although the inferred physical separation is compatible with the diffusion radius range.

Our results suggest instead that the gamma-ray emission from J1930+1826E may more likely be related to PWN/SNR G54.1+0.3, but a deeper investigation is warranted to secure the association, in particular to verify that the required gas density ($n_{\rm H}\sim20-30~{\rm cm^{-3}}$) holds over the full extent of the source.

In an alternative interpretation, the source J1930+1826E may be connected to a young and massive stellar cluster (YMSC). Indeed, J1930+1826E is spatially coincident (with an offset $0\fdg09$) with the optical cluster Teutsch 42 (DSH J1930.2+1832), which is part of a young open cluster population ($<$30 Myr) characterized by high mechanical wind power \citep{Celli2024}. The interaction of supersonic stellar winds with the cluster environment is expected to create strong shock waves and turbulent conditions, propitious for particle acceleration, possibly at the highest energies \citep{Bykov2013, Morlino2021}. Indeed, GeV and TeV emissions have been detected in the directions of some YMSCs \citep{Abramowski2012,Aharonian2019,Cao2024SciBu}. This scenario certainly deserves further investigation.

\subsection{Origin of J1929+1732E}

Source J1929+1732E is extended with a 68\% containment radius of $0\fdg45 \pm 0\fdg04$ (assuming a 2D Gaussian morphology).
Figure \ref{fig:tsmap1} shows that this extension encompasses several known high-energy objects or sources, including the gamma-ray pulsar PSR J1928+1746, radio pulsar PSR J1928+1725, the TeV source HAWC J1928 (offset $\sim0\fdg39$), and the UHE source 1LHAASO J1928+1746u (offset $\sim0\fdg29$). Because of the existence of that HAWC source in positional coincidence with PSR J1928+1746, and since this pulsar has a relatively high spin-down power, we favored a scenario in which all sources are ultimately related to PSR J1928+1746, with nonthermal particles energized by the PWN and/or the SNR. 
In that respect, the different source extents (from $r_{68} = 0\fdg45$ in the LAT range down to $r_{39} \simeq 0\fdg18$ at HAWC and $r_{39} \simeq 0\fdg17$ at LHAASO energies) and their relative positional offsets might indicate that we are probing different populations of sources and/or different radiation processes. The properties of PSR J1928+1746 are $P = 69$ ms, $\dot{E}=1.6\times10^{36}~\mathrm{erg~s^{-1}}$, and $\tau_c\simeq8.2\times10^{4}~\mathrm{yr}$ \citep{Cordes2006}, which a priori provides comfortable energy and time budgets for particle acceleration and propagation.

From our model fits, the GeV$-$TeV gamma-ray emission from J1929+1732E and HAWC J1928 cannot be adequately described by a single leptonic or hadronic model.
The GeV emission shows a distinct spectral break at low energies, resembling that seen in evolved SNRs interacting with molecular clouds such as IC 443 and W44 \citep{Ackermann2013Sci}, where the gamma rays arise from the PD process.
On the other hand, the flattening of the spectrum at TeV energies, and the smaller sizes of HAWC J1928 and 1LHAASO J1928+1746u, could indicate that an additional component of a different nature dominates increasingly at the highest energies.
This combination of GeV and TeV features motivated a two-component hadronic–leptonic model. For the particle distribution of each component, an ECPL model was adopted. 
This combined model nicely reproduces the spectra observed by \textit{Fermi}-LAT and HAWC (see bottom-right panel of Figure \ref{fig:naima_sed} and Table \ref{tab:sed modeling}). We provide separate discussions below for each component.

\subsubsection{Origin of the GeV emission}

\citet{Albert2023ApJ} analyzed the $^{13}$CO molecular cloud and found three velocity peaks of approximately 4.5, 22, and 46 km s$^{-1}$ in the region of HAWC J1928. Using the strongest component at 22 km s$^{-1}$, they derived an ambient gas density of $n_{\rm H}=50~\mathrm{cm^{-3}}$ (see their Table 5). 
%
To explain the GeV emission using a hadronic model, we adopted this density estimate as a starting point and a distance of $D=4.3~\mathrm{kpc}$ to PSR J1928+1746 \citep{Cordes2006}. 
The fitting of our model shows good consistency with the GeV spectrum, which exhibits a steep rise below $\sim1$~GeV and a turnover around a few GeV, consistent with the characteristic $\pi^0$ bump expected from hadronic gamma-ray production \citep{Ackermann2013Sci,Peron2020ApJL,Tibet2021NatAs}.
The resulting total energy in protons above 100 MeV is $W_p=3.1\times10^{49}~\mathrm{erg}$, which falls well within the canonical fraction ($\sim5\%-10\%$) of the kinetic energy of a typical supernova explosion that can be converted into CRs via diffusive shock acceleration. One natural candidate here is the supernova that gave birth to PSR J1928+1746, which may have become undetectable because of the large age of the system ($\tau_c\simeq8.2\times10^{4}~\mathrm{yr}$).

We note that our best-fit proton index ($\Gamma \approx 1.95$) is consistent, within uncertainties, with the canonical diffusive shock acceleration prediction of 2.0, albeit slightly harder. This slight hardening can be attributed to the reacceleration of preexisting Galactic CRs in an evolved SNR environment, as observed in the Cygnus Loop \citep{Tutone2021}. Additionally, the use of kinetic energy rather than momentum in our numerical modeling may further contribute to this spectral hardening in the transrelativistic regime ($<10$ GeV).

An important caveat is that the assumed gas density was determined in \citet{Albert2023ApJ} for a relatively small region (two gas clumps of 12 and 18 pc in diameter), much smaller than the physical extent of the LAT source (about $1^\circ$ at 4 kpc or 70 pc). 
So, as in the case of J1930+1826E, it remains to be verified that a relatively high average gas density can be found over the extent of J1929+1732E. In the opposite case, for instance, more typical ISM gas densities $n_{\rm H}=1-10~\mathrm{cm^{-3}}$, the energy required for CR protons to power the observed GeV emission would exceed what a single supernova can provide, and an alternative explanation should be envisioned.

\subsubsection{Origin of the TeV emission}

The TeV emission of HAWC J1928 can be well described by a leptonic component arising mainly from IC scattering of relativistic electrons injected by PSR J1928+1746 and accelerated in its putative PWN.
The total energy in relativistic electrons above 100 MeV was derived as $W_{e}~(\mathrm{>100\, MeV}) \simeq 0.4 \times10^{48}$ erg, while the energy in relativistic electrons above 1 TeV is $W_{e}~(\mathrm{>1\, TeV}) \simeq 3.8\times10^{46}$ erg, which is consistent with the value of $W_{e}\approx4.6\times10^{46}$ erg obtained by \citet{Albert2023ApJ} from independent modeling of the HAWC data only. 
Assuming a spherical emission region with a volume of $V \sim 3.7\times10^{4}~\mathrm{pc^{3}}$ \citep[Table 4 in][]{Albert2023ApJ}, the corresponding mean electron energy density is $\epsilon_{\mathrm{e}}= W_{e}/V \simeq 0.20~\mathrm{eV~cm^{-3}}$. 
%
This value is smaller than the characteristic energy density of the ISM ($\epsilon_{\mathrm{ISM}} \simeq 1~\mathrm{eV~cm^{-3}}$). 
Given the age of PSR J1928+1746, such a low value of $\epsilon$ is consistent with an old, diffuse PWN, in which the relativistic electrons have started to cool and diffused into a large volume over time. 
This is consistent with the pulsar halo definition proposed in \citet{Giacinti2020AA}.

Under the assumption of a small initial spin-down timescale (i.e., $\tau_0 = 10$–$100$ yr), the maximum possible rotational energy released by PSR J1928+1746 is given by
$E_{\mathrm{tot}} = \dot{E}_0 \tau_0 \simeq (3.4\times10^{51}$–$3.4\times10^{52})~\mathrm{erg}$,
where $\dot{E}_0 \simeq (1.1\times10^{42}–1.1\times10^{44})~\mathrm{erg~s^{-1}}$ is the initial spin-down luminosity derived from Eq.~\ref{eq:cal_E_dot} with the current age of the pulsar estimated as $t = \tau_c - \tau_0 \simeq (8.19$–$8.20)\times10^4~\mathrm{yr}$. 
Thus, the lower limit of the energy-conversion efficiency is $\eta = W_{e} / E_{\mathrm{tot}}\simeq 0.001\%–0.01\%$. This value is significantly lower than the efficiencies inferred for PWNe. For instance, in the Crab Nebula, this efficiency is estimated to be as high as 30\% \citep[e.g.,][]{Amato2024arXiv}, and up to nearly 100\% in many other PWNe \citep{Torres2014JHEAp}. This would tend to suggest that the initial spin-down timescale is not as small as assumed in the above calculation and lies rather in the range of a few tens of thousands of years.
On the other hand, as mentioned before, this did not take energy losses into account since particles were injected, possibly tens of thousands of years ago ($\tau_c\simeq8.2\times10^{4}~\mathrm{yr}$ for PSR J1928+1746).

Overall, the low electron energy density, extended nature of HAWC J1928 and 1LHAASO J1928+1746u, and old age of the PSR J1928+1746 all suggest that HAWC J1928 could be an old PWN, where relativistic electrons have propagated away and cooled over time, consistent with the properties of extended TeV PWNe reported by the H.E.S.S. Collaboration \citep{HESS_Collaboration2018A2}. 
Our analysis is fully consistent with the results of \citet{Albert2023ApJ} and further strengthens the interpretation that the TeV emission of HAWC J1928 originates from an evolved PWN or TeV halo \citep{Jardin2022} powered by PSR J1928+1746, though currently no PWN has been firmly identified in this region. 


\section{J1925+1729P, a likely new pulsar?}
\label{newPSR}

As revealed by the spatial analysis, the gamma-ray source J1925+1729P is significantly offset from the radio position of PSR J1925+1720 even after taking the systematics into account in position determination. Besides, the broadband emission of J1925+1729P shows a curved spectrum turning down above 1 GeV, which is reminiscent of pulsars detected by the \textit{Fermi}-LAT \citep{3PC}, although the LP model is slightly preferred over the PLEC4 model with a $\Delta \log \mathcal{L}\sim3$. We speculate that there are two distinct sources within this small region around J1925+1729P: one located at the position of PSR J1925+1720 and the other at J1925+1729P. 

Assuming that J1925+1729P is a new yet undetected pulsar and following the approach outlined in 3PC, we can determine the peak energy $E_{p}$ of the SED and the curvature at $E_{p}$ given by 
\begin{equation}
E_{p}=E_{0}\left[(1+\frac{b}{d}(2-\Gamma)\right]^{\frac{1}{b}} \, \, ,
\end{equation}
and
\begin{equation}
d_{p}=d+b(2-\Gamma)  \, \, .
\end{equation}
In this formulation, $d_{p}$ attains a maximum value of $4/3$ for synchrotron or curvature radiation produced by monoenergetic electrons. 
The width of the SED peak at $E_{p}$ is inversely correlated to $d_{p}$. A higher value of $d_{p}$ corresponds to a narrower peak, indicating that the emission is dominated by electrons within a relatively narrow energy range. Conversely, a lower $d_{p}$ results in a broader peak, reflecting contributions from a wider span of electron energies. 
In the spectral analysis, modeling J1925+1729P with a PLEC4 model ($b$ was fixed to the canonical value of 2/3) resulted in $\Gamma=2.14$ and $d=0.83$. We then obtained $E_{p}=0.84$ GeV and $d_{p}=0.74$ for J1925+1729P, locating it near the upper-left edge of the most crowded part of the $d_{p}$ vs. $E_{p}$ plot (see Figure 20 in the 3PC). Following 3PC and given the relatively low value of $d_{p}$, we suggest that the emission detected from the new pulsar candidate J1925+1729P is produced from a rather broad electron population. 

Interestingly, J1925+1729P is spatially coincident (positional offset of $0\fdg20$) with cluster G052.073+00.713 \citep[or G52.04+0.55,][]{Urquhart2014}, which hosts massive young stellar objects and HII regions. It is therefore reasonable to speculate that J1925+1729P is a newborn pulsar in this star-forming region. 
Firmly pinpointing the pulsar nature of J1925+1729P requires the detection of pulsations from radio, X-ray, or gamma-ray observations. Failing to detect pulsations might suggest that the very engine powering the source could be another object in the star-forming region, or maybe the star-forming region itself, although in the latter case, one may expect some extension depending on the nature of the emitting particles and the exact mechanisms at play in their acceleration and transport.

\section{Summary and conclusions} 
\label{conclusion}
In this work, we performed a detailed \textit{Fermi}-LAT analysis to characterize the gamma-ray emission properties in a region that encompasses three 1LHAASO sources in the Galactic plane within longitudes $52^{\circ}<l<55^{\circ}$. This region is best modeled by three pointlike sources and two extended sources, which provides an improved representation compared to the one in the \textit{Fermi}-LAT 4FGL-DR4 catalog. Broadband gamma-ray theoretical modeling, especially in combination with HAWC data, was performed using the \texttt{naima} package under different emission scenarios. We summarize our findings below.

\begin{enumerate}

\item Source J1932+1916P is most likely the gamma-ray pulsar PSR J1932+1916, whose emission in the LAT band is predominantly of magnetospheric origin. Its TeV counterpart, HAWC J1932, likely represents the associated PWN.

\item Source J1925+1729P exhibits a pulsar-like spectrum, yet is spatially inconsistent with the known gamma-ray pulsar PSR J1925+1720. This suggests that it is a new gamma-ray pulsar candidate, pending confirmation via dedicated pulsation searches.

\item Source J1930+1851P coincides with the composite PWN/SNR system G54.1+0.3 and the TeV source HAWC J1930. The joint LAT and HAWC spectra can be interpreted by either leptonic (PWN) or hadronic (SNR) scenarios, possibly involving escaped particles illuminating nearby gas in the latter case.

\item Extended source J1930+1826E shows no obvious extended TeV counterpart. From the spectral and energetic standpoint, a connection with PWN/SNR G54.1+0.3 is viable, by which J1930+1826E would be powered by escaped particles from the system. The relation to the overlapping source 1LHAASO J1929+1846u needs to be clarified, which may involve a revised modeling of the Galactic diffuse emission in the LAT and/or LHAASO bands.

\item Extended source J1929+1732E is spatially associated with PSR J1928+1746, the TeV sources HAWC J1928 and 1LHAASO J1928+1746u. Joint LAT–HAWC modeling favors a hybrid lepto-hadronic origin in which GeV emission arises from hadronic interactions, possibly CRs escaped from an SNR and interacting with nearby molecular clouds, and TeV emission results mainly from a leptonic process, consistent with an old PWN or TeV halo powered by PSR J1928+1746.

\end{enumerate}

In conclusion, our refined \textit{Fermi}-LAT analysis provides a comprehensive spatial and spectral characterization of a complex Galactic region hosting multiple pulsars, SNRs, and TeV sources. Joint theoretical modeling of \textit{Fermi}-LAT and HAWC (whenever available) data for three sources detected in this work gives a refined broadband overview of this complex region, beyond the work done in \cite{Albert2023ApJ}. Additional HAWC and LHAASO observations will be useful to further constrain the emission mechanisms and strengthen the proposed associations. In a more distant future, CTAO observations with their improved angular resolution should bring a crucial piece of information to the study of the region, with implications on topics as diverse as the formation of pulsar halos or particle escape from SNRs.

\begin{acknowledgments}
We thank David A. Smith and Jean Ballet for the very useful comments and suggestions that helped to improve the article. 
This work is supported by the National Key R\&D Program of China (2023YFE0101200). We acknowledge supports from the National Natural Science Foundation of China under grant No. 12373051 and 12233006. 
The \textit{Fermi}-LAT Collaboration acknowledges generous ongoing support from a number of agencies and institutes that have supported both the development and the operation of the LAT, as well as scientific data analysis. These include the National Aeronautics and Space Administration and the Department of Energy in the United States; the Commissariat \`a l'Energie Atomique and the Centre National de la Recherche Scientifique/Institut National de Physique Nucl\'eaire et de Physique des Particules in France; the Agenzia Spaziale Italiana and the Istituto Nazionale di Fisica Nucleare in Italy; the Ministry of Education, Culture, Sports, Science and Technology (MEXT), High Energy Accelerator Research Organization (KEK), and Japan Aerospace Exploration Agency (JAXA) in Japan; and the K.~A.~Wallenberg Foundation, the Swedish Research Council, and the Swedish National Space Board in Sweden. Additional support for scientific analysis during the operations phase is gratefully acknowledged from the Istituto Nazionale di Astrofisica in Italy and the Centre National d'\'Etudes Spatiales in France. This work was performed in part under DOE contract DE-AC02-76SF00515.
\end{acknowledgments}

%

\vspace{5mm}
\facility{\textit{Fermi}}
\software{\textit{naima}}







\bibliography{bibtex}{}
\bibliographystyle{aasjournal}



\end{document}